\documentclass[a4paper,11pt]{article}
\usepackage[english]{babel}
\usepackage{graphicx,amsmath,amsfonts,amssymb,slashed,upgreek,color,caption,amscd,cite,cancel,footnote}
\usepackage{multirow}
\usepackage{subfigure}
\DeclareGraphicsExtensions{.pdf,.png,.jpg}

\usepackage{amsbsy}
\usepackage{empheq}
\usepackage{hyperref}
\usepackage{pstricks}

\newcommand{\rhom}{\rho_{\rm m}}
\newcommand{\Om}{\Omega_{\rm m}}
\newcommand{\Omo}{\Omega_{\rm m,0}}
\newcommand{\Ol}{\Omega_{\rm DE}}

\newcommand{\wo}{w_{0}}
\newcommand{\wa}{w_{a}}

\newcommand{\boldg}{\boldsymbol{\gamma}}
\newcommand{\eos}{$EoS$ }
\newcommand{\x}{\omega}

\newcommand{\MNRAS}{MNRAS }
\newcommand{\AANDA}{A\&A }
\newcommand{\PR}{Phys.~Rev. }
\newcommand{\PRL}{Phys.~Rev.~Lett. }

\newcommand{\APJ}{ApJ }
\newcommand{\APJS}{ApJS }
\newcommand{\AJ}{AJ }
\newcommand{\IJMP}{Int.~J.~Mod.~Phys. }
\newcommand{\IJTP}{Int.~J.~Theor.~Phys. }
\newcommand{\JCAP}{JCAP }

\begin{document}

\begin{center}

\Large{\textbf{Probing non-standard gravity  with  the growth index: \\  a background independent analysis}} \\[1cm]
\large{Heinrich Steigerwald$^{\rm a}$, Julien Bel$^{\rm b}$ and Christian Marinoni$^{\rm a,c}$}
\\[0.5cm]

\small{
\textit{$^{\rm a}$ Aix Marseille Universit\'e, Universit\'e de Toulon, CNRS, CPT UMR 7332, 13288, Marseille, France \\
}}

\small{
\textit{$^{\rm b}$ INAF - Osservatorio Astronomico di Brera, Via Brera 28, Milano, via Bianchi 46, 23807 Merate, Italy\\
}}

\small{
\textit{$^{\rm c}$  Institut Universitaire de France, 103, bd. Saint-Michel, F-75005 Paris, France
}}

\vspace{.2cm}

\end{center}

\vspace{2cm}

\begin{abstract}

Measurements of the growth index of linear matter  density fluctuations  $\gamma(z)$ provide a clue as to whether Einstein's field equations encompass gravity also on large cosmic scales, those where 
the expansion of the  universe accelerates. We show that the  information encoded in this function can be satisfactorily parameterized using   a small set of coefficients $\gamma_i$, 
in such a way that the true scaling of the growth index is recovered  to better than  $1\%$ in most dark energy and dark  gravity models. 
We find that the  likelihood  of  current  data,  given this formalism and  the  $\Lambda$ Cold Dark Matter ($\Lambda$CDM) expansion model of $Planck$, is maximal for $\gamma_0=0.74^{+0.44}_{-0.41}$ and $\gamma_1=0.01^{+0.46}_{-0.46}$, a measurement  compatible with the $\Lambda$CDM predictions ($\gamma_0=0.545$,  $\gamma_1=-0.007$). 
In addition,   data tend to  favor models  predicting slightly less growth of structures than the $Planck$  $\Lambda $CDM scenario. 
The main aim of the paper is to  provide a prescription for routinely calculating, in an analytic way,  the amplitude of the growth indices $\gamma_i$ in relevant cosmological scenarios, and to show  that these parameters naturally define a space where   predictions of alternative theories of gravity can be compared against growth data in a manner which is independent from the expansion history of the cosmological background.  As the standard $\Omega$-plane provides a tool to identify different expansion histories $H(t)$ and their relation to various  cosmological models,  the $\gamma$-plane can thus be used to locate different growth rate histories $f(t)$ and their relation to alternatives model of gravity. As a result,  we find that the Dvali-Gabadadze-Porrati gravity model is rejected with a $95\%$ confidence level. By simulating  future data sets,    such as those that a Euclid-like mission will provide,  we also show how to tell apart $\Lambda$CDM  predictions from those of more extreme possibilities,  such as smooth dark energy models, clustering quintessence or parameterized post-Friedmann cosmological models.
\end{abstract}

\newpage 
\tableofcontents

\section{Introduction}
\label{sec:introduction}
Measurements of the  expansion rate history $H(t)$ of  the universe, when interpreted within the 
standard model of cosmology, convincingly indicate
that the universe has recently entered a phase of accelerated expansion \cite{Perlmutter:1998np,Riess:1998cb, astier, marpairs,mod1, anderson, Bel:2013ksa, san,Ade:2013zuv}. 
Most of this unexpected evidence  is provided via
geometric probes of cosmology, that is  by constraining  the redshift scaling  of the luminosity distances $d_L(z)$  
of cosmic standard candles (such as Supernovae Ia), or of the angular diameter distance $d_A(z)$ of 
cosmic standard rulers (such as the sound horizon scale at the last scattering epoch).

Despite much observational evidence, little is known about  the physical
mechanism that drives cosmic acceleration. As a matter of fact,  virtually all the attempts to make sense of  this perplexing  
phenomenon without invoking a new constant  of nature (the so called cosmological constant)  call for  exotic physics beyond current theories.
For example, it is speculated that cosmic acceleration might be
induced by a non clustering,  non evolving,   non interacting and extremely light 
vacuum energy $\Lambda$  \cite{Peebles:2002gy},
or by a cosmic field with negative pressure, and thus repulsive
gravitational effect, that changes with time and
varies across space (the so called dark energy fluid) \cite{lucashin, cope, wett:1988, Caldwell:1997ii, ArmendarizPicon:2000dh,Binetruy:2000mh,Uzan:1999ch,Riazuelo:2001mg, Gasperini:2001pc}, if not by a break-down of Einstein's
theory of gravity on cosmological scales (the so called dark gravity scenario) \cite{costas, DeFelice:2010aj, DGP, Deffayet:2001pu, Arkani-Hamed:2002fu,Capozziello:2003tk,  Nojiri:2005jg, deRham:2010kj,  Piazza:2009bp,GPV,JLPV,BFS}.

This last, extreme eventuality is made somewhat less far-fetched  by the fact that 
a large variety of nonstandard gravitational models,  inspired by fundamental physics arguments,
can be finely tuned  to reproduce the expansion rate history of the successful  standard model of cosmology, the 
$\Lambda$CDM paradigm. 
Although different models make undistinguishable predictions about the amplitude and scaling of background observables such as $d_L, d_A$ and $H,$ the analysis of the  clustering properties of matter on large linear cosmic scales is in principle sufficient to distinguish and falsify  alternative gravitational scenarios. Indeed, a generic prediction of  
modified gravity theories is that the Newton constant $G$ becomes a time (and possibly scale) dependent function $G_{\rm eff}$. Therefore, dynamical observables of cosmology 
which are sensitive to the amplitude of $G$,  such as,  
for example,  the  clustering  properties of cosmic structures,  provide a probe for  resolving geometrical  degeneracies among models  and for properly identifying  the specific signature of nonstandard  gravitational signals. Considerable phenomenological effort is thus devoted to engineering  and applying methods for 
extracting information from dynamical observables of  the inhomogeneous sector of the universe 
\cite{Bel:2012ya,TurHudFel12,DavNusMas11,BeuBlaCol12,PerBurHea04,SonPer09,RosAngSha07,CabGaz09,SamPerRac12, ReiSamWhi12,ConBlaPoo13,GuzPieMen08,TorGuzPea13}.
Indeed, thanks to large and deep future galaxy redshift surveys, such as for example Euclid \cite{euclid2},  
the clustering  properties of matter  will be soon characterized   with a `background level'  precision,  
thus providing us with  stringent constraints on the  viability of alternative  gravitational scenarios. 

Extending  the perimeter of precision cosmology beyond zeroth order observables into the domain of 
first order perturbative quantities critically depends  on observational improvements but also
on  the refinement  of theoretical tools. Among the quantities that are instrumental in constraining 
modified gravity models, the linear growth index $\gamma$, 

\begin{equation} \label{defuno}
\gamma(a) \equiv \big( \ln \Omega_{\rm m} (a) \big)^{-1} \ln \Big( \frac{d \ln \delta_{\rm m}(a)}{d \ln a} \Big)
\end{equation}

\noindent where $a$ is the scale factor of the universe, $\Omega_{\rm m}=(8\pi G \rho_{\rm m})/(3H^2)$ is  the reduced density of matter and $\delta_{\rm m} =\rho_{\rm m}/\bar{\rho}_{\rm m} -1$ 
the dimensionless density contrast of matter, has attracted much attention. Despite being in principle a function, this quantity is often, and effectively,  parameterized as being  constant 
\cite{pee80}. Among the various appealing properties of such an approximation,  two in particular deserve to be mentioned. 
First, the salient features of the growth rate history of linear structures   can be  compressed into a single scalar quantity which can be easily constrained using standard parameter estimation 
techniques. As it is the case with parameters such as $H_0$, $\Omega_{\rm m,0}$, etc.,  which incorporate  all the information contained in the expansion rate  function $H(t)$, so it is 
extremely  economic to label and identify different growth  histories $\delta_{\rm m}(t)$ with the single book-keeping index  $\gamma$.
Moreover, since the growth index parameter takes distinctive values for distinct gravity theories, any deviation of its estimated amplitude  from  the reference value  $\gamma_0=6/11$ (which represents the 
exact asymptotic early value of the function $\gamma(a)$ in a $\Lambda$CDM cosmology \cite{WanSte98})
is generically interpreted as a potential signature of new gravitational physics.

However useful in quantifying deviations from standard gravity predictions, this index   must also be precise to be  of any practical use.
As a rule of thumb, the systematic error introduced by approximating  $\gamma(a)$ with  $\gamma_0$, which  depends on $\Om$, must be much smaller 
than the precision with which future experiments are expected to constrain the growth index  over a wide redshift range($\sim 0.7\%$ \cite{euclid2}).  
Notwithstanding, already within  a standard $\Lambda$CDM framework with $\Omega_{\rm m,0}=0.315$, the imprecision of the  asymptotic  approximation 
 is of order $2\%$ at $z=0$.  More subtly, the expansion kinematic is expected to leave time dependent  imprints in the growth index. 
  The need to model the redshift evolution of the growth index, especially in view of the large  redshift baseline that will be surveyed by  future data, 
led to the development of  more elaborated parameterizations \cite{Gong, GanMorPol09, FuWuYu2009,pg,gp}. Whether their justification is  purely phenomenological or 
theoretical,  these  formulas aim at locking the expected  time variation of $\gamma(a)$ into a small set  of scalar quantities, the so called growth index parameters  $\gamma_i$. 
For example,  some authors (e.g. \cite{pg,gp}) suggest to use the Taylor expansion $\gamma(z)\,=\,\gamma_0\,+\,\big[\frac{d \gamma}{dz}\big]_{z=0}\, z$ for data fitting purposes.  Indeed, this approach 
has the merit of  great accuracy at present epoch, but it becomes too inaccurate at the  intermediate redshifts  ($z\sim 0.5$) already probed by current data.

On top of precision issues,  there are also interpretational concerns. Ideally, we would like the growth index parameter space to be also in one-to-one correspondence with predictions of specific gravitational theories. In other terms we would like to use likelihood contours in this plane to select/reject specific gravitational scenarios. This is indeed   a tricky issue.  
For example,  it is  rather involved  to link the amplitude  of the growth index parameters to predictions of  theoretical models if the growth index fitting formula has a phenomenological nature. 
More importantly,   it is not evident how to extract growth information ($\delta_{\rm m}(a)$) from a function, $\gamma$  which, as equation (\ref{defuno}) shows,  is degenerate with background information (specifically  $\Om(a)$). In other terms, the growth index parameters are model dependent quantity that can be estimated  only after  a specific model for the evolution of the background quantity $\Om(a)$
is supplied. Therefore it is not straightforward to use the likelihood function in the $\gamma$-plane to reject dark energy scenarios   for which  the background quantities do not scale as in the fiducial.
Because of this, up to now,  growth index measurements in a given fiducial were used  to rule out only  the null-hypothesis that the fiducial  correctly describes  large scale structure formation processes. Growth index estimates  were not used to gauge the viability   of a  larger class of alternative gravitational scenarios.  
 
A reverse argument also holds and highlights the importance of working out a 
growth index parameterization which is able to capture the finest details of the  exact numerical solution, establishing at the same time, the functional dependence on background observables.
Indeed, once a given gravitational paradigm is assumed as a prior,  the degeneracy of growth index measurements with background information,  can be exploited to constraining the 
background parameter of the resulting  cosmological model, directly using  growth data. Therefore, by expressing the  growth index  as a function of  specific dark energy or dark gravity parameters one can 
test for the overall coherence of  cosmological information extracted from the joint analyses of  smooth and inhomogeneous observables.

In this paper we  address some of these issues  by means of  a new parameterization of the growth index. The main virtues of the approach is that the proposed formula
is   {\it  a)}  flexible, i.e.~it describes   predictions of a wide class of cosmic acceleration models, {\it b)} accurate,  i.e.~it performs better than alternative models  in reproducing exact numerical results,  {\it c)} it is 'observer friendly',    i.e.~accuracy is  achieved with a minimum number of parameters
 and  {\it d)} it is  `theorist friendly', i.e.~the amplitude of the fitting parameters can be directly and mechanically related, in an analytic way, to predictions of theoretical models.

The paper is organized as follows. We define the parameterization for the growth index in section \S \ref{sec:parametrizing}, and we discuss its  accuracy in describing various dark energy models  such as smooth  and clustering quintessence in  \S \ref{sec:standard}. In \S \ref{sec:modified}  we apply the formalism to modified gravity scenarios. In particular,  we discuss the DGP \cite{DGP} and the Parameterized Post  Friedmanian \cite{FerSko10} scenarios. In \S \ref{sec:constraining} we will impose the studied models to current (simulated future) data.  Conclusions are drawn in \S \ref{sec:conclusions}. Throughout all the paper, if not specified differently, the  flat Friedmann-Lema\^itre-Robertson-Walker cosmology with $Planck$ parameters $\Omo =0.315, \sigma_{8,0}=0.835$ \cite{Ade:2013zuv} is referred to as the {\it reference} $\Lambda$CDM model.

\section{Parameterizing the growth index}
\label{sec:parametrizing}
In this section we introduce our notation and  we give the theoretical background for analyzing the clustering of matter on large linear scales. 
For a large class of dark energy models and gravitational laws, at least on  scales where the quasi-static approximation applies, 
the dynamics of linear matter perturbations $\delta_{\rm m}$ can be effectively described by the  following second order differential equation
\begin{align}
\ddot{\delta}_{\rm m} + 2\:\! H \:\! \nu \:\! \dot{\delta}_{\rm m} - 4\:\! \pi \:\! G \:\! \mu \:\! \rho_{\rm m} \:\! \delta_{\rm m} = 0,
\label{eq:matter_density_fluctuations}
\end{align}

\noindent where overdot ($\,\dot{}\,$) denotes derivation with respect to cosmic time $t$ and 
the dimensionless response $\mu \equiv G_{\rm eff}/G$ and  damping  $\nu$ coefficients are general functions of cosmic time and possibly the spatial Fourier frequency.  As for the simplest case, General Relativity even augmented by a minimally coupled  scalar field,  marks the point $\mu=\nu=1$ while for modified gravity models we expect, in general,  $\mu  \neq 1$.
The specific form  of $\mu$ predicted  in higher dimension brane models \cite{DGP} or in scalar tensor theories of gravity \cite{FerSko10}  are given in \S \ref{sec:DGP} and \S \ref{sec:skofe} respectively.
See, instead, \cite{kun} for a more elaborate model of modified gravity which does not reduce to the standard form of Eq. \ref{eq:matter_density_fluctuations}.

It is convenient, from the observational perspective,  to express Eq. \ref{eq:matter_density_fluctuations} 
in terms of the linear  growth rate $f$,   a cosmic observable defined as the logarithmic
derivative of the matter density fluctuations $f=\frac{d\ln \delta_{\rm m}}{d \ln a}$. We obtain  
\begin{align}
f'+f^2+\Big(1+\nu +\frac{H'}{H}\Big)\:\! f - \frac{3}{2}\!\: \mu\, \Omega_{\rm m} = 0
\label{eq:f_H_General}
\end{align}
where prime ($'$) denotes derivation with respect to $\ln a$. It is standard practice   to look for solutions of Eq. \ref{eq:f_H_General} with the form 
\begin{align}
f=\Omega_{\rm m}(a)^{\,\gamma(a)}
\label{eq:f_approximation}
\end{align}
where  $\gamma$ is called the growth index already introduced in \S \ref{sec:introduction}.  Although, in some special cases, an exact analytic solution can be  
explicitly given (for example,  in terms of a hypergeometric function if  $\nu=\mu=1$ and the DE {\it EoS} is constant  \cite{sil}), 
the approximate parametric solution $\gamma=constant$  has been widely advocated \cite{sp,kn,ius,pg,sa,nep,gp,cdp1,bbs,cdp2,car,basil}. Indeed,  
the growth rate of matter density perturbations  is highly sensitive to the kinematics of the  smooth expansion of the Universe, 
as well as to the gravitational instability properties of the background fluids. The approximate anstaz $\gamma=constant$ interestingly and   
neatly  splits the background (basis) and the inhomogeneous  contribution (exponent)
to the linear growth rate of structures. Gravitational models  beyond  standard general relativity, 
which predict identical background expansion histories, i.e.~identical observables $H(a)$ and  $\Omega_{\rm m}(a)$,
may thus be singularized and differentiated by constraining the amplitude of  $\gamma$. 
However solutions with a constant growth index of (sub-horizon) matter density perturbations 
cannot be realized in quintessence models of dark energy over  the whole period from the beginning of  matter domination up into distant future.
Moreover, it is likely that different gravitational models tuned to reproduce the same expansion history $H$
display unidentical evolution for $\Omega_{\rm m}$. This implies that  the amplitudes of  $\gamma$ inferred in two different cosmological models, characterized by 
distinctive evolution laws  of the matter density parameter,  cannot be directly compared. 

In order to increase the accuracy of predictions, we look for  a more elaborate parametric form of the growth index: we express $\gamma$ as a function of $\ln \Omega_{\rm m}$. Plugging (\ref{eq:f_approximation}) into (\ref{eq:f_H_General}) we obtain
\begin{align}
\x '\:\! \Big(\gamma(\x ) + \x \:\! \frac{d\gamma}{d\x }\Big) + e^{\x \:\! \gamma(\x )} + 1 + \nu(\x ) + \frac{H'}{H}(\x ) - \frac{3}{2}\!\; \mu(\x )\;\! e^{\x (1-\gamma(\x ))} = 0
\label{eq:f_H_x}
\end{align}
where we set  $\x =\ln \Omega_{\rm m}$. We assume that all non-constant coefficients ($\x '$, $\tfrac{H'}{H}$, $\mu$, $\nu$) are well-defined functions of $\x $, and are
completely specified once a  theory of gravity  is considered. Since the numerical solution of equation (\ref{eq:f_H_x}) for a $\Lambda$CDM model 
indicates that $\gamma$ is an almost linear function of $\x $, and guessing  that viable  alternative theories of gravity 
should predict  minimal, but detectable,  deviations from standard model results, we suggest  to describe  the time evolution of the
growth index as

\begin{align}
\gamma(\x ) = \sum_{n=0}^{N} \gamma_n \frac{\x ^n}{n!} + \mathcal{O}(\x ^{N\:\!\! +\:\!\! 1})
\label{eq:gamma_Taylor}
\end{align}
where $\{ \gamma_0,\gamma_1,\ldots,\gamma_N\}$ are constant parameters.  Expressing $\gamma$ via a series expansion has already been proposed  earlier by \cite{WanSte98} and improved later by 
\cite{Gong}. In those works $\x = 1-\Omega_{\rm m}$, whereas here we set  $\x =\ln \Omega_{\rm m}$. Demonstrating the gain in accuracy achieved by this change of variable 
is one of the goals of this paper.  A different one, but equally important,  is to show that by this choice we can work out  a closed  analytic formula that predicts  the amplitude of the coefficients $\gamma_n$ (up to an arbitrary order $N$)  once any given dark energy and gravitational model is specified. Specifically, we define the  structural parameters  of the formalism
\begin{align}\label{eq:struct}
\mathcal{X}_n := \big[\tfrac{d^n}{d\x ^n}\big(\x '\big)\big]_{\x =0} \;, \quad \mathcal{H}_{n} := \big[\tfrac{d^n}{d\x ^n}\big(\tfrac{H'}{H}\big)\big]_{\x =0} \;,  \quad \mathcal{M}_{n} := \big[\tfrac{d^n}{d\x ^n}\big(\mu \big)\big]_{\x =0} \;, \quad \mathcal{N}_{n} := \big[\tfrac{d^n}{d\x ^n}\big(\nu \big)\big]_{\x =0} 
\end{align}
where $n$ is a natural number and $\tfrac{d^0}{d\x ^0} \equiv 1$. We obtain (see \S \ref{sec:Annexe})
\begin{align}
\gamma_0 = \frac{3\:\!(\mathcal{M}_0 +\mathcal{M}_{1}) - 2\:\!( \mathcal{H}_{1} + \mathcal{N}_{1})}{2 + 2\:\! \mathcal{X}_{1} + 3 \:\! \mathcal{M}_0}
\label{eq:gamma_0} 
\end{align}
For $n \geq 1$, we have
\begin{align}
\gamma_n =& \:\! 3 \frac{\mathcal{M}_{n\:\!\! +\:\!\! 1}\:\! +\:\! \sum_{k=1}^{n\:\!\! +\:\!\! 1}\:\!\! \binom  {n\:\!\! +\:\!\! 1}{k} \:\! \mathcal{M}_{n\:\!\! +\:\!\! 1\:\!\! -k}\:\!  B_{k}\:\!\!(1-y_1,-y_2,-y_3,\ldots ,-y_k)}{(n+1) \big(2+2\:\!(n+1)\mathcal{X}_1 + 3\:\! \mathcal{M}_0 \big)} \nonumber  \\
   &- 2 \frac{ \:\! B_{n\:\!\! +\:\!\! 1}\:\!\!(y_1,y_2,\ldots ,y_{n\:\!\! +\:\!\! 1})\:\!  + \:\! \sum_{k=2}^{n\:\!\! +\:\!\! 1}\:\!\! \binom{n\:\!\! +\:\!\! 1}{k}\:\!  \mathcal{X}_{k} \:\! (n+2-k)\:\! \gamma_{n\:\!\! +\:\!\! 1\:\!\! -\:\!\! k}\:\!  + \:\! \mathcal{H}_{n\:\!\! +\:\!\! 1}  + \:\! \mathcal{N}_{n\:\!\! +\:\!\! 1} }  {(n+1) \big(2+2\:\!(n+1)\mathcal{X}_1 + 3\:\! \mathcal{M}_0 \big)}
\label{eq:gamma_n}
\end{align}
where 
\begin{align}
y_i=
\begin{cases}
i\:\! \gamma_{i-1} & \text{if }i \leq n \\
0 & \text{if } i=n+1
\end{cases}
\end{align}
and the Bell polynomials are defined by
\begin{align}
B_k(x_1,x_2,\ldots ,x_k)=\sum \tfrac{k!}{j_1! j_2!\cdots j_k!} \big(\tfrac{x_1}{1!}\big)^{j_1} \big(\tfrac{x_2}{2!}\big)^{j_2}\cdots \big(\tfrac{x_k}{k!}\big)^{j_k}
\end{align}
where the sum is taken over all $k$-tupels $\{j_1,j_2,\ldots ,j_k\}$ of non-negative integers satisfying
\begin{align}
1 \times j_1+2 \times j_2 + \ldots + k\times j_k = k.
\end{align}
In the following we will test the precision of the approximation (\ref{eq:gamma_Taylor}) for different accelerating models and for various cosmological parameters. In particular we will show that two coefficients $\gamma_0$ and $\gamma_1$ are sufficient for a large class of models.

\section{Standard Gravity}
\label{sec:standard}
In this section we    frame our analysis in a  Friedmann Lema\^{i}tre Roberston \& Walker (FLRW) model of the universe, a metric model characterized by  two degrees of freedom: the spatial curvature of the universe  $k$ and the scale factor $a(t)$. We therefore assume that gravity is ruled by the standard Einstein field equations, and we further assume that the hyper-surfaces of constant time are flat ($k=0$) as  observations  consistently suggest \cite{Ade:2013zuv}, and that, at least during the late stage of its evolution, i.e.~the period we are concerned with, the universe is only filled with cosmic matter and
dark energy. These are perfect fluid components whose $EoS$, i.e.~the ratio  $w(a)=p(a)/\rho(a)$ between pressure and density, is zero and lower than $-1/3$ respectively. We refer to the particular case in which the DE $EoS$ $w=-1$ as to the $\Lambda$CDM model. The evolution of the physical density of matter and DE  is given by 
\begin{equation}
\rho(a)=\rho_0 \;\! a^{-3(1+\tilde{w}(a))}
\label{eq:conservation}
\end{equation}
\noindent   where

\begin{equation}
\tilde{w}(a)=\frac{1}{\ln(a)}\int_{1}^{a} w(a') \;\! d\ln a',
\end{equation}

\noindent while the expansion rate of the cosmic metric,  or simply the Hubble parameter $H(a)$,  is given by 

\begin{equation}
H^2(a)=H_0^2\left[ \Omega_{\rm m,0} \;\! a^{-3}+(1-\Omega_{\rm m,0}) \;\! a^{-3(1+\tilde{w}(a))}\right]
\label{eq:Friedmann}
\end{equation}

\noindent where $\Omega_{\rm m,0}$ is the present day reduced density of matter. In this notation, the evolution equations for the reduced density of matter and DE are 
\begin{equation}
\Omega_{\rm m}(a)=\left ( 1+\frac{ 1-\Omo}{\Omo}\;\! a^{-3\tilde{w}(a)} \right )^{-1}
\label{eq:omegam}
\end{equation}
\noindent and
\begin{equation}
\Omega_{\rm DE}(a)=1-\Omega_{\rm m}(a)
\end{equation}
\noindent respectively. 

Under these hypotheses,   the relevant quantities in Eq. (\ref{eq:f_H_x})  become
\begin{align}
\x '(\x ) \equiv \tfrac{d \ln \Omega_{\rm m}}{d \ln a} = 3 \big(1-\Omega_{\rm m}(a) \big) w(a)  \;, \quad \tfrac{H'}{H}(\x ) \equiv \tfrac{d\ln H}{d\ln a} = -\tfrac{3}{2}\big(1+(1-\Omega_{\rm m}(a))w(a)\big)
\label{eq:relations_SG}
\end{align}
where $w(a)$ is the (possibly varying)  $EoS$ parameter  of the dark energy fluid. We will illustrate the performances of the proposed parameterization (cf. Eq.  (\ref{eq:gamma_Taylor}))
first by assuming that dark energy is a smooth component which does not cluster on any scale, i.e.~its energy density is a function of time only.  We will then increase the degrees of freedom associated to 
this hypothetical component,  by assuming that  dark energy may become gravitationally  unstable, that is its effective density varies also as a function of space.

\subsection{Smooth Dark Energy}
\label{sec:smooth}

The propagation of instabilities in the energy density and pressure of a fluid can be heuristically understood
in terms of two kinematical quantities: the  speed of sound  $c^2_s=\delta p/\delta \rho$ and the sound horizon $L_s=c_s/H$.
On scales larger than the sound horizon instabilities can collapse. On scales below $L_s$  the fluid is smooth.
As an example of this last case, one can consider a canonical, minimally coupled, scalar field such as quintessence. 
Since  $c_s=1$,  the sound horizon coincides with  the cosmic horizon 
and the quintessence fluid can be considered homogeneous on all  (observationally) relevant  scales. 
In our notation,  such a smooth dark energy component is effectively described by setting 
\begin{align}
\mu(\x )=1  \quad \text{and} \quad \nu(\x ) =1. 
\label{eq:relations_SDE}
\end{align}
into Eq. (\ref{eq:f_H_x}). Setting
\begin{align}
\mathcal{W}_n := \big[\tfrac{d^n}{d\x ^n}\big(w(a)\big)\big]_{\x =0}\;
\label{eq:gamma_coeff_w}
\end{align}
we  find
\begin{align}
\mathcal{X}_0 = 0 \;, \quad \mathcal{X}_n = -3\:\! \sum_{i=0}^{n-1}\binom {n}{i} \mathcal{W}_i \;, \quad \mathcal{H}_0 = -\frac{3}{2} \;, \quad \mathcal{H}_n = -\frac{1}{2} \mathcal{X}_n = \frac{3}{2} \sum_{i=0}^{n-1}\binom {n}{i} \mathcal{W}_i \;,
\label{eq:gamma_coeff_x_h}
\end{align}
for $n=1,2,3,$ etc. Smooth models of dark energy  make physical sense only for $w \ge -1$. This {\it null energy condition}, indeed,  prevents both ghost and gradient instabilities (e.g. \cite{PSM}). Notwithstanding, 
sound theories of DE can be constructed also by relaxing the null energy condition,  for example, by imposing the Lagrangian density of the gravitational field to contain
space derivatives higher than two, i.e.~terms that become important at some high momentum scale (see  \S 2.2 of Ref.~\cite{Creminelli:2008wc}). 
In what follows,  we will thus consider also  super acceleration scenarios $w<-1$, their effective, phenomenological character being  understood.

We now discuss  two distinct scenarios. First,  we  assume  that the dark energy {\it EoS} varies weakly with cosmic time, i.e.~that it can be effectively described in terms
of a constant parameter $w$. Then we complete our analysis by computing the explicit amplitude of the growth index parameters in a  scenario in which the DE {\it EoS} is free to vary.
\begin{figure}[t] \vspace{-1cm}
\centering
\includegraphics[trim = 0cm 0cm 0cm 0cm, scale=0.39]{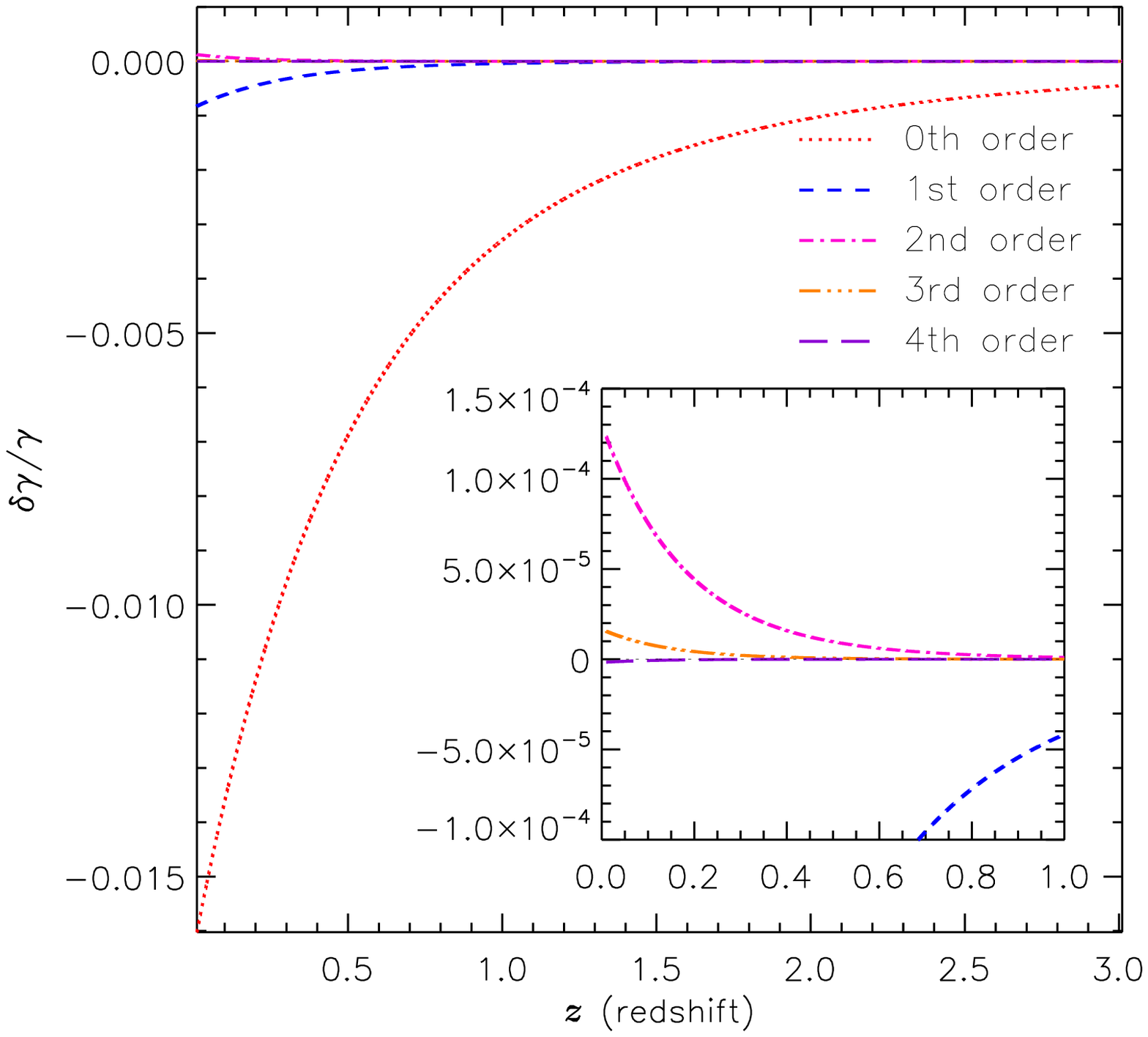}
\includegraphics[trim = 0cm 0cm 0cm 0cm, scale=0.39]{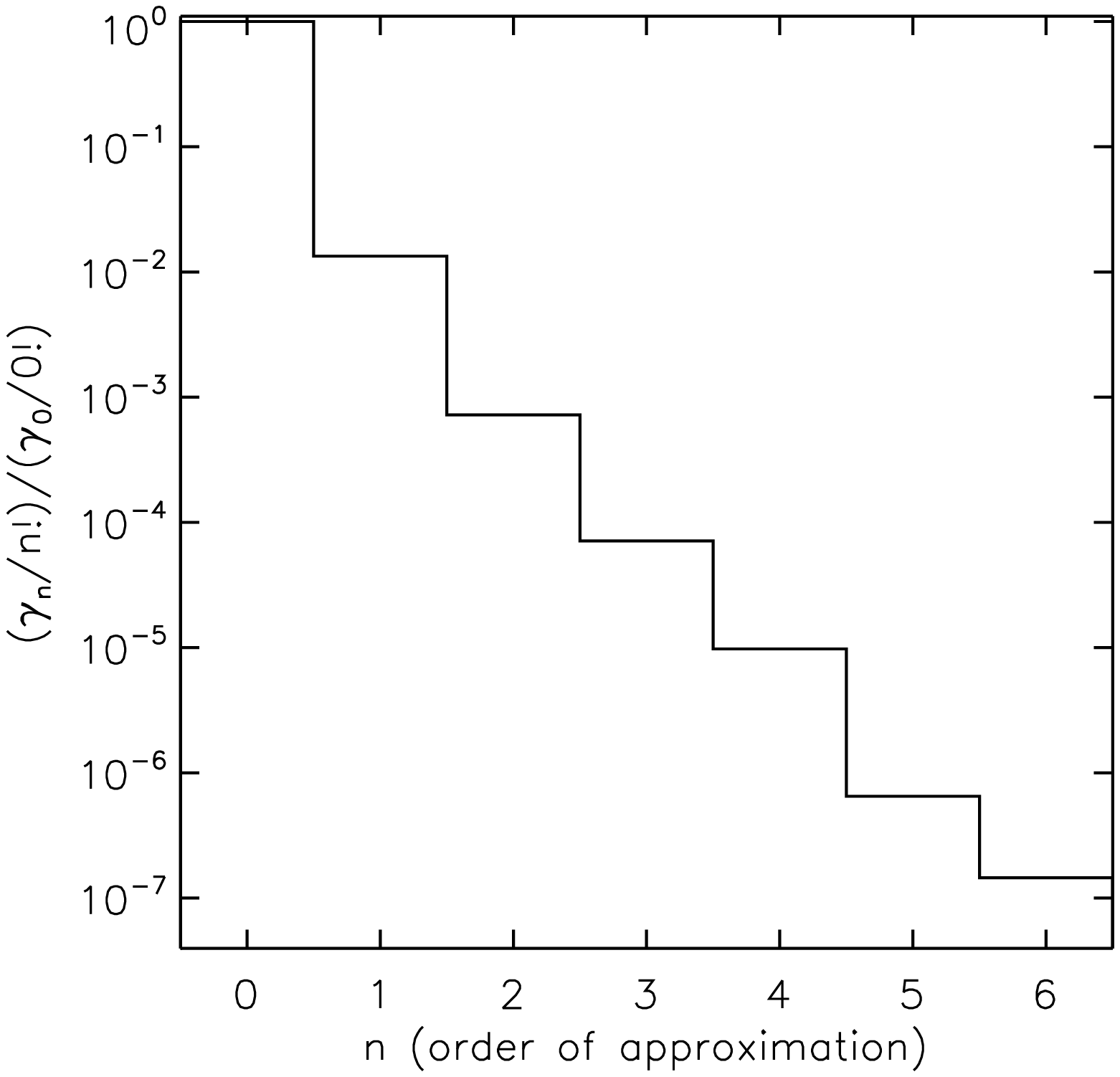}
\includegraphics[trim = 0cm 0cm 0cm 0cm, scale=0.39]{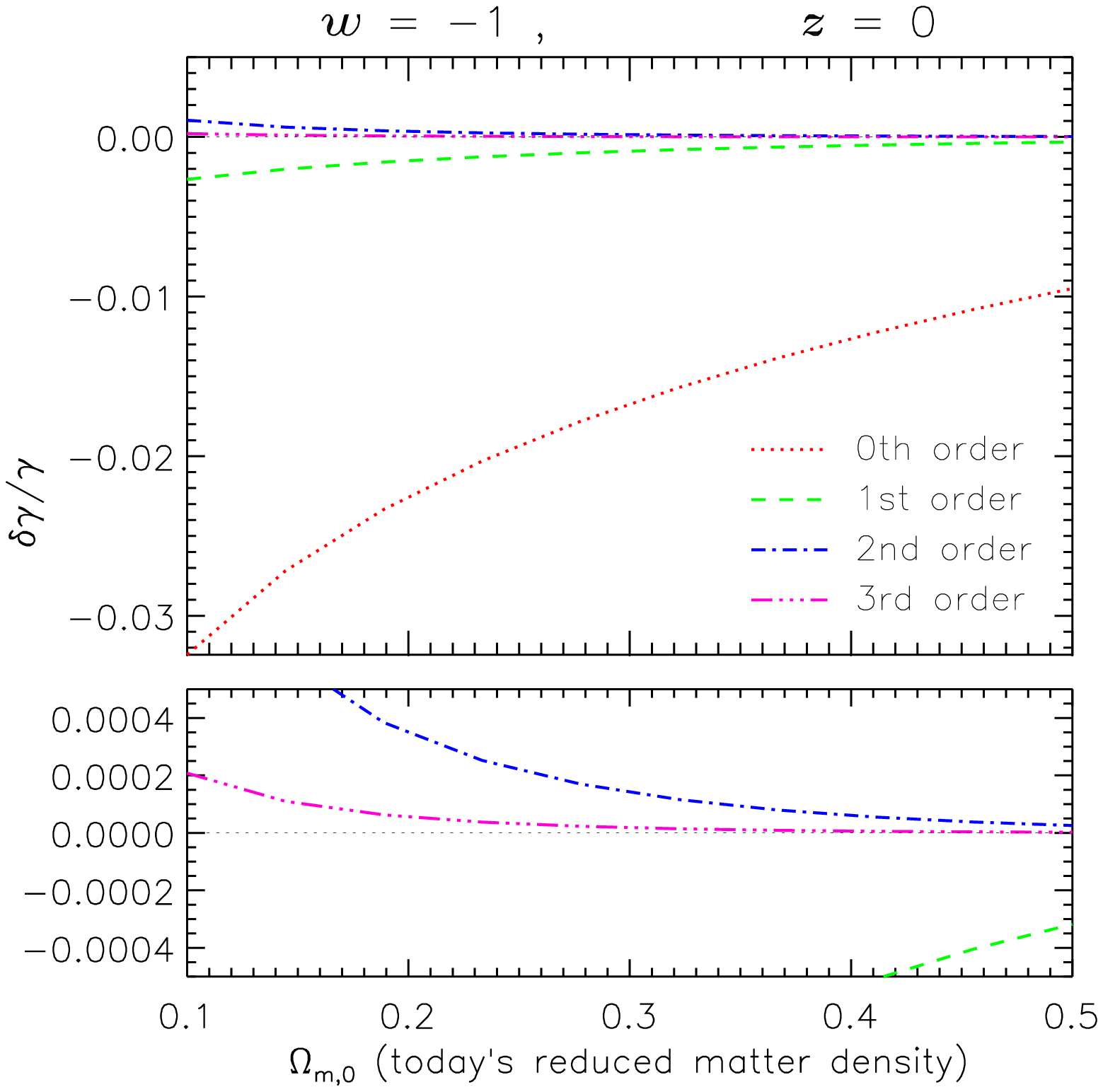}
\includegraphics[trim = 0cm 0cm 0cm 0cm, scale=0.39]{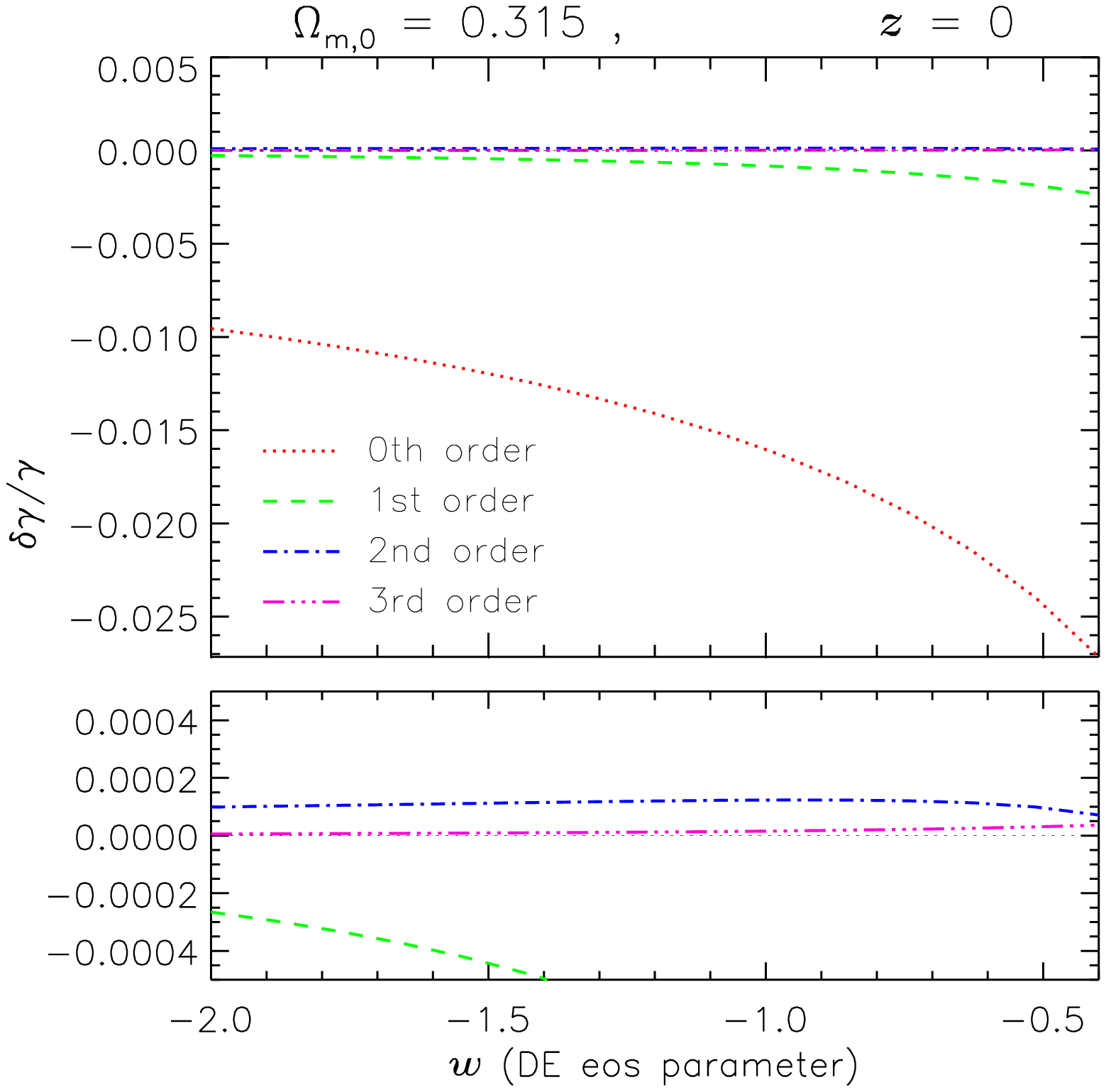}
\caption{{\it Upper  left panel:} relative imprecision of the approximation (\ref{eq:gamma_Taylor}) for the growth index $\gamma$ in a smooth Dark Energy model with constant equation of state $w$  as a function of redshift   and of the expansion order. The exact value of the growth index is obtained by solving numerically Eq. (\ref{eq:f_H_General}) and by using Eq. (\ref{defuno}).
{\it Upper right panel:} amplitude of  the Taylor series coefficients of (\ref{eq:gamma_Taylor}) (normalized to $\gamma_0$)
as a function of the expansion order (up to 6). The relative accuracy increases  roughly by one order of magnitude going from one order to the next higher one. {\it Lower left panel:} relative imprecision at $z=0$ as a function of $\Omega_{\rm m,0}$ in a $\Lambda$CDM model  showing the stability of the approximation within  a sensible range of matter density values. {\it Lower right panel:} relative imprecision as a function of DE equation of state parameter $w$.}
\label{fig:wCDM}
\end{figure}

\begin{figure}[t]
\centering
\includegraphics[scale=0.7]{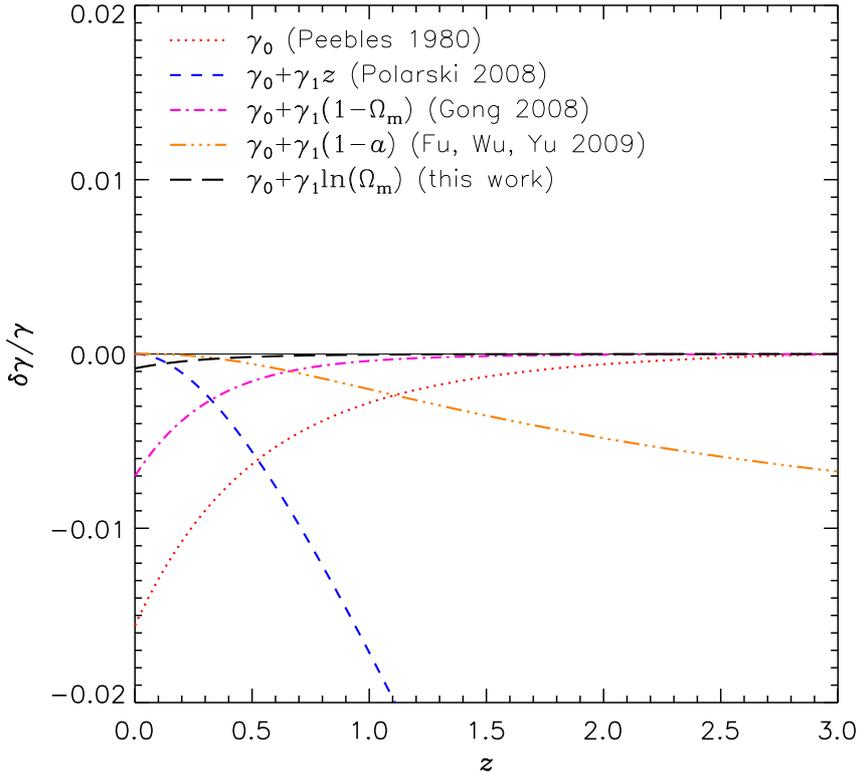}
\caption{The precision of various parameterizations\cite{pee80,pg,Gong,FuWuYu2009} of the growth index is shown for  the  {\it reference} $\Lambda$CDM  model. 
The precision is estimated as the relative difference with respect  to  the numerical reconstruction of the growth index $\gamma(z)$ in the reference model.}
\label{fig:comparison}
\end{figure}

\subsubsection{DE with a constant equation of state $w$}\label{sec:w}
In this case, in Eq.~\ref{eq:gamma_coeff_w} we have $\mathcal{W}_0=w$ and  $\mathcal{W}_n=0$ for $n \geq 1$.  We therefore obtain from Eq. (\ref{eq:gamma_coeff_x_h})

\begin{align}
\mathcal{X}_0 = 0 \;, \quad \mathcal{X}_{n} = -3\:\! w \;,\quad \mathcal{H}_0 = -\tfrac{3}{2} \;, \quad \mathcal{H}_{n} = \tfrac{3}{2} \:\! w \;, \quad \mathcal{M}_{0} = \mathcal{N}_0 = 1\;,\quad  \mathcal{M}_{n} = \mathcal{N}_{n} = 0 \;,  \label{eq:coeff_GR_1}
\end{align}
where $n=1,2,3,$ etc. Replacing (\ref{eq:coeff_GR_1}) in (\ref{eq:gamma_0}) and (\ref{eq:gamma_n}) we find:
\begin{align}
\gamma_0 = \tfrac{3\:\! (1-w)}{5-6\:\! w} \;, \quad \gamma_1 = -\tfrac{3\:\!(1-w)(2-3\:\! w)}{2\:\!(5-12\:\! w)(5-6\:\! w)^2}  \;,\quad \gamma_2 = \tfrac{(1-w)(485-3\:\! w\:\!(1015-3\:\! w\:\! (559-270\:\! w\:\! )))}{10\:\! (5-18\:\! w)(5-12\:\! w)(5-6\:\! w)^3} 
\label{eq:gamma012wCDM}
\end{align}
In particular, for the $\Lambda$CDM model (i.~e.~for $w=-1$), we find 
\begin{align}
\gamma_0^{\Lambda \rm CDM} = \frac{6}{11} \simeq 0.54545 \;,\quad
\gamma_1^{\Lambda \rm CDM} = -\frac{15}{2\:\! 057} \simeq -0.00729 \;, \quad
\gamma_2^{\Lambda \rm CDM} = \frac{410}{520\:\! 421} \simeq 0.00079
\label{eq:gamma012LambdaCDM}
\end{align}

For $\Omega_{\rm m,0}=0.315$, the relative error $\delta\gamma/\gamma < 0.1\%$ at order 1, $\delta\gamma/\gamma < 0.02\%$ at order 2 and $\delta\gamma/\gamma < 0.002\%$ at order 3.
More generally,  Figure \ref{fig:wCDM} (upper right panel) shows that the accuracy of the approximation  increases  by roughly a factor of 10 as soon as a new coefficient is switched on. 
This same figure (lower right panel) shows  that the relative accuracy depends only slightly on the equation of state parameter $w$, although the error is in general slightly larger for larger $w$. 
The relative error of the approximation is also quite stable as we vary $\Omega_{\rm m,0}$ 
within a sensitive range of  values $\Omega_{\rm m,0} = [0.2,0.4]$ (see the lower left panel in  Figure \ref{fig:wCDM}).

We remark that the coefficients $\gamma_0$ and $\gamma_1$ are the same\footnote{up to the sign of $\gamma_1$} when expanding the growth index in powers of $(1-\Omega_{\rm m})$ whereas $\gamma_2$ and higher orders differ. However, as a consequence of developing in $\ln \Omega_{\rm m}$,  already at first order our approximation is, depending on cosmology, from  5 to 10 times better at $z=0$. 
This is seen in Figure~\ref{fig:comparison},  where we also compare the accuracy of Eq.~\eqref{eq:gamma012wCDM} with various other parameterizations of the growth index. 
Not only Eq.~\eqref{eq:gamma_Taylor} is more precise than formulas
based on perturbative expansions around $z=\infty$, in the relevant redshift range for dark energy studies, that is $z>0.5$, it is also more accurate than the phenomenological \cite{Fu Wu Yu 2009} or perturbative \cite{pg} expressions normalized at $z=0$.

\begin{figure}[t]
\centering
\includegraphics[scale=0.333]{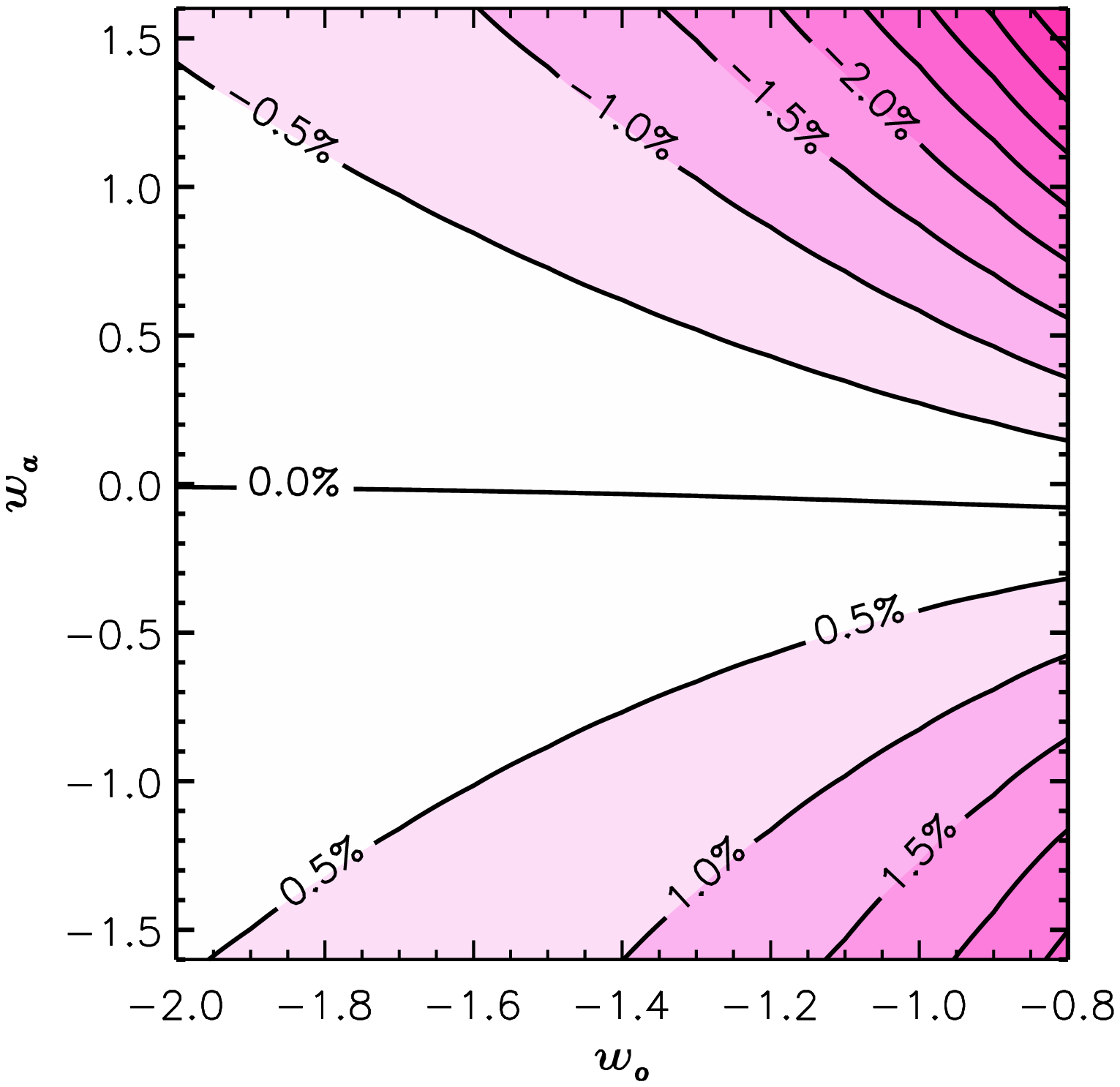}
\includegraphics[scale=0.333]{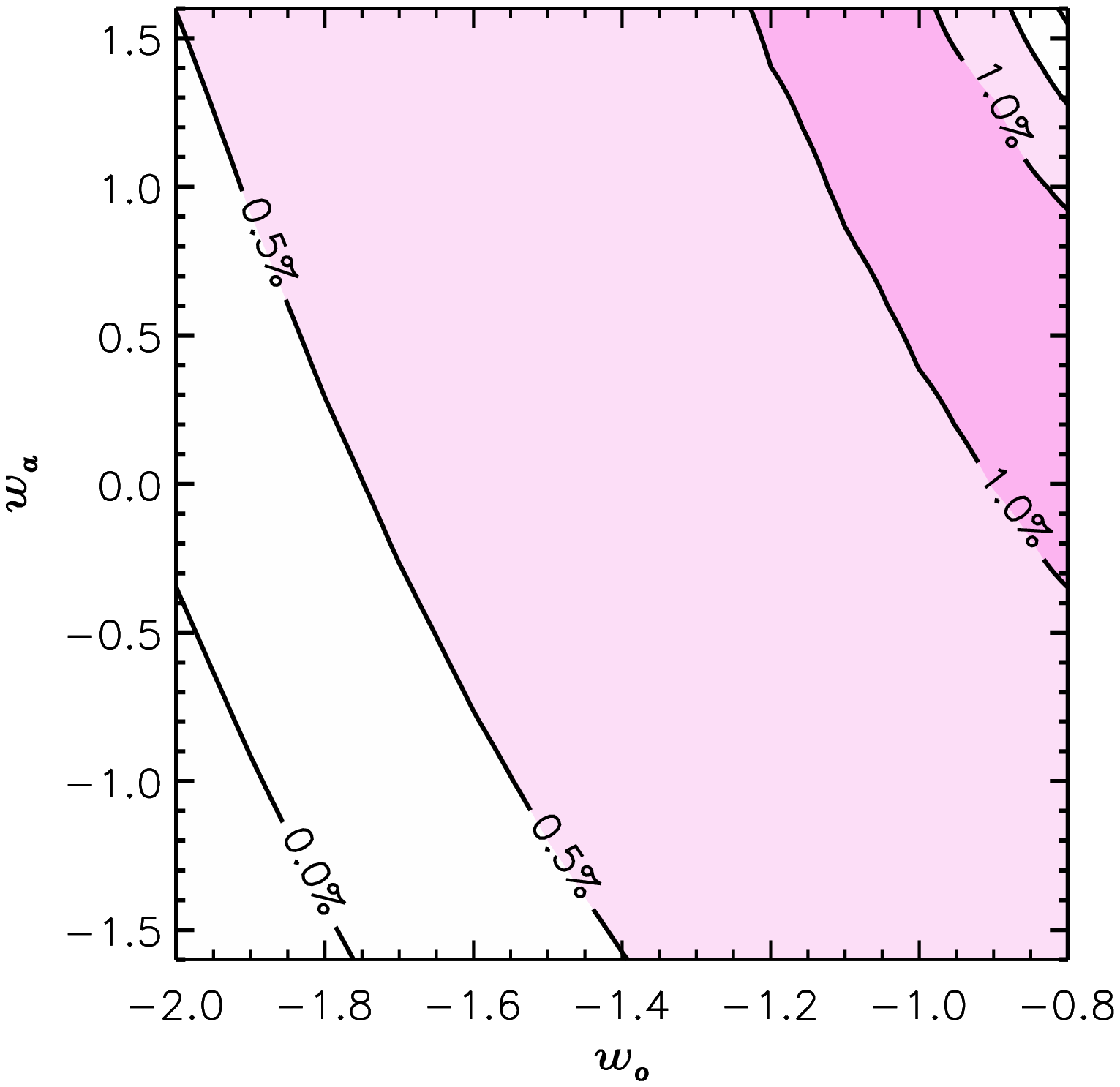}
\includegraphics[scale=0.333]{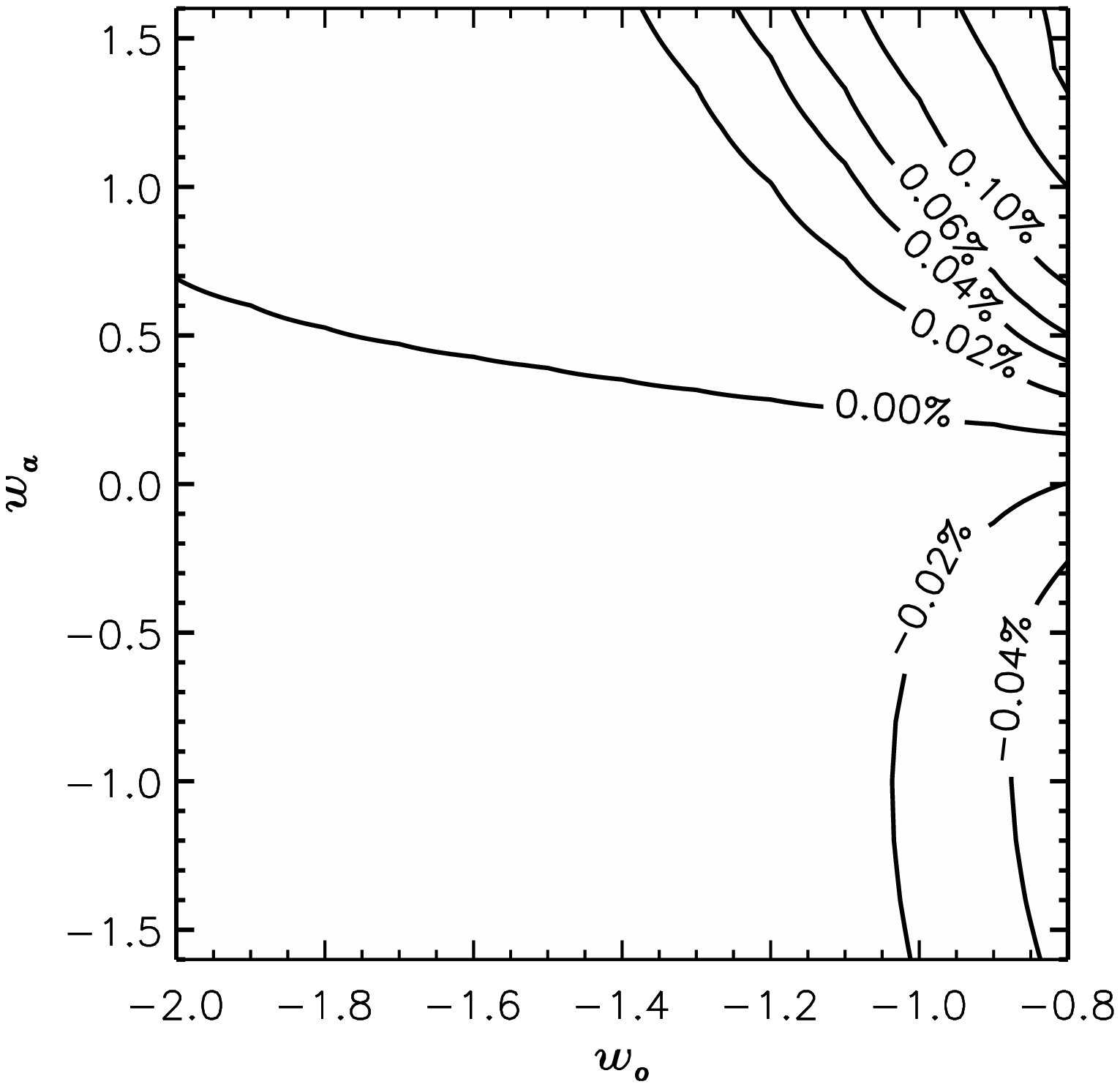}
\caption{{\it Left panel:} Relative error, as a function of the  $EoS$ parameters ($w_{\rm o}, w_{a} $), in approximating the growth index of  smooth dark energy  models
with  the first order model $\gamma_0+\gamma_1(1-\Omega_{\rm m})$  of \cite{WanSte98}.  {\it Middle panel}: Accuracy of the phenomenological fitting formula  of \cite{lin}. {\it Right panel}: Accuracy of the first order growth index model $\gamma_0+\gamma_1 \ln \Omega_{\rm m}$. In all plots we use the same color scale and we assume a  flat cosmological model with $\Omega_{\rm m}=0.315$. The relative error represents the maximal imprecision in the  redshift range surveyed by a Euclid-like survey ($0.7<z<2$).  }
\label{fig:w(a)CDM_a}
\end{figure}

\subsubsection{Varying equation of state $w(a)$}
\label{sec:w(a)}
We first compute the coefficients  $\mathcal{W}_n$ (see Eq. \ref{eq:gamma_coeff_w}). 
If we assume $w_i \equiv w(a=0) < - \frac{1}{3}$, then the limit $\x =\ln \Omega \rightarrow 0$ is equivalent to the limit $ a \rightarrow 0$.  We obtain 
\begin{align}
\label{eqs:coeff_w}
\mathcal{W}_0 = \big[w(a)\big]_{a=0}\,,\;
\mathcal{W}_1 = \big[(\x ')^{-1} w'\big]_{a=0} \,,\;
\mathcal{W}_2 = \big[(\x ')^{-2}\big(3w'w-(w')^2/w+w''\big)-(\x ')^{-1} w'\big]_{a= 0} \,, 
\end{align}
etc., where $\x '= 3w(1-e^{\x })$ from equation (\ref{eq:relations_SG}). 
The formalism requires the knowledge of the $n^{th}$ order derivative of the DE {\it EoS} $w(a)$  for $a \rightarrow 0$, that is well beyond the redshift domain where  the expansion rate of the universe is expected to be sensitive to contributions from the dark energy density.  Noting that $\x ' \rightarrow 0$ when $a \rightarrow 0$, the finiteness of the coefficients $\mathcal{W}_i$ (at least up to $\mathcal{W}_2$) can be enforced by requiring the {\it EoS} to satisfy the following equation 
\begin{align}
3\:\! w'\:\! w - (w')^2/w + w'' = 0 \label{eq:condition_W_finite}
\end{align}
whose most general solution is
\begin{align}
w(a) = w_i \Big(1 - \frac{\wa}{3 \wo^2}a^{-3w_i}\Big)^{-1} \;, \quad
w_i =  \wo\Big(1-\frac{\wa}{3\wo^2}\Big)
\label{eq:wi_finite}
\end{align}
where $\wo=w(a=1)$, $\wa=-\big[\tfrac{dw}{da}\big]_{a=1}$.  Note that  by linearizing the previous expression around $a=1$ one recovers the  standard  parameterization $w(a) = \wo + \wa (1-a)$ \cite{Chevallier Polarski 2000}. For this specific equation of state, we have
\begin{align}
\displaystyle
\mathcal{W}_0= \wo \Big(1-\frac{\wa}{3\wo^2}\Big)^{-1} \;, \quad \mathcal{W}_n = (-1)^n \frac{\wa \Omo}{3\wo (1-\Omo)} \; \text{for} \; n>0
\label{eq:Wi_finite}
\end{align}
and, as a consequence,   the coefficients of the series expansion which defines the growth index  are 
\begin{subequations}\label{eq:gammanw(a)CDM}
\begin{align}
\gamma_0 &= \tfrac{3(1-w_i)}{5-6 w_i} \label{eq:gamma0w(a)CDM}\\
\gamma_1 &= -\tfrac{3(1-w_i)(2 - 3 w_i)-6(5-6w_i)\mathcal{W}_1}{2(5-12w_i)(5-6w_i)^2}\label{eq:gamma1w(a)CDM}\\
\gamma_2 &= \tfrac{(1-w_i)(2-3w_i)(11-30w_i)-3(5-6w_i)(23-6w_i(10-6w_i))\mathcal{W}_1 -72(5-6w_i)^2\mathcal{W}_1^2+3(5-12w_i)(5-6w_i)^2\,\mathcal{W}_2}{(5-12w_i)(5-18w_i)(5-6w_i)^3}\label{eq:gamma2w(a)CDM}.
\end{align}
\end{subequations}

The relative variation of the growth index $\gamma(z)$ with respect to the $\Lambda$CDM value, when the {\it EoS} parameters  span  the range   ($-2<\wo<-0.8, -1.6<\wa<1$)
can be as large as $3.7\%$ at $z=0$. As a reference, if  $\wo$ is fixed to the value $-1$, then one can expect a relative difference as important as   $1.5\%$, 
a figure larger  than the nominal precision with which a Euclid-like survey is expected to constrain $\gamma$, that is $0.7\%$ \cite{euclid2}.
Figure \ref{fig:w(a)CDM_a} shows that within the  redshift range that will be surveyed by Euclid ($0.7<z<2$), already when truncating the expansion at first order, the maximal 
imprecision with which  the redshift scaling of  $\gamma$  is reconstructed over all the parameter space  is  less than 0.1\%. 

We are not aware of any  theoretically justified model that accounts for possible variability in the DE {\it EoS}, so we compare the accuracy of our parameterization  (cf. Eqs.~\eqref{eq:gammanw(a)CDM})  with  that  obtained using the Wang \& Steinhardt (1998) model\cite{WanSte98}, which is supposed to be an accurate approximation also for 
weakly varying dark energy equation of states.  Additionally, we also consider the phenomenological fitting  formula of \cite{lin},  
($\gamma\,=\, 0.55\,+\,[0.05\, \Theta(w+1)\,+\,0.02\, \Theta(1+w)\,]\,(1\,+\,w(z=1)\,)$,  where $\Theta(x)$ is the Heaviside step function 
and $w$ is the constant  equation of state ({\it EoS}) parameter of the dark energy fluid). 
As can be appreciated by inspecting Figure \ref{fig:w(a)CDM_a}, the gain in precision with respect to both these growth index models can be as high as 1 order of magnitude 
in the range    ($-2<\wo<-0.8, -1.6<\wa<1$),  a figure which pushes the systematic error  well below the expected measurement precision of the Euclid satellite.

\begin{figure}[t]
\centering
\includegraphics[scale=0.47]{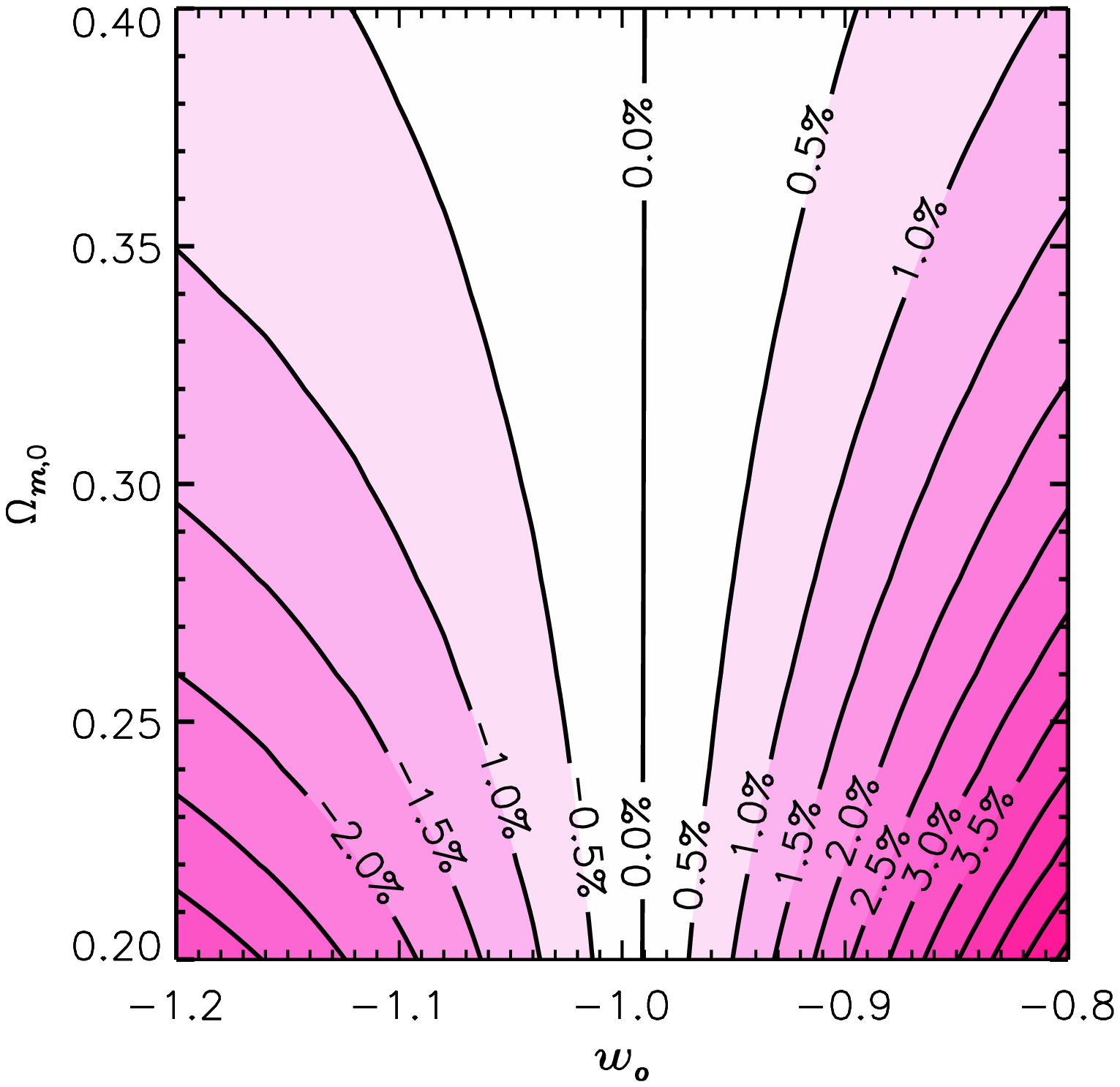}
\includegraphics[scale=0.47]{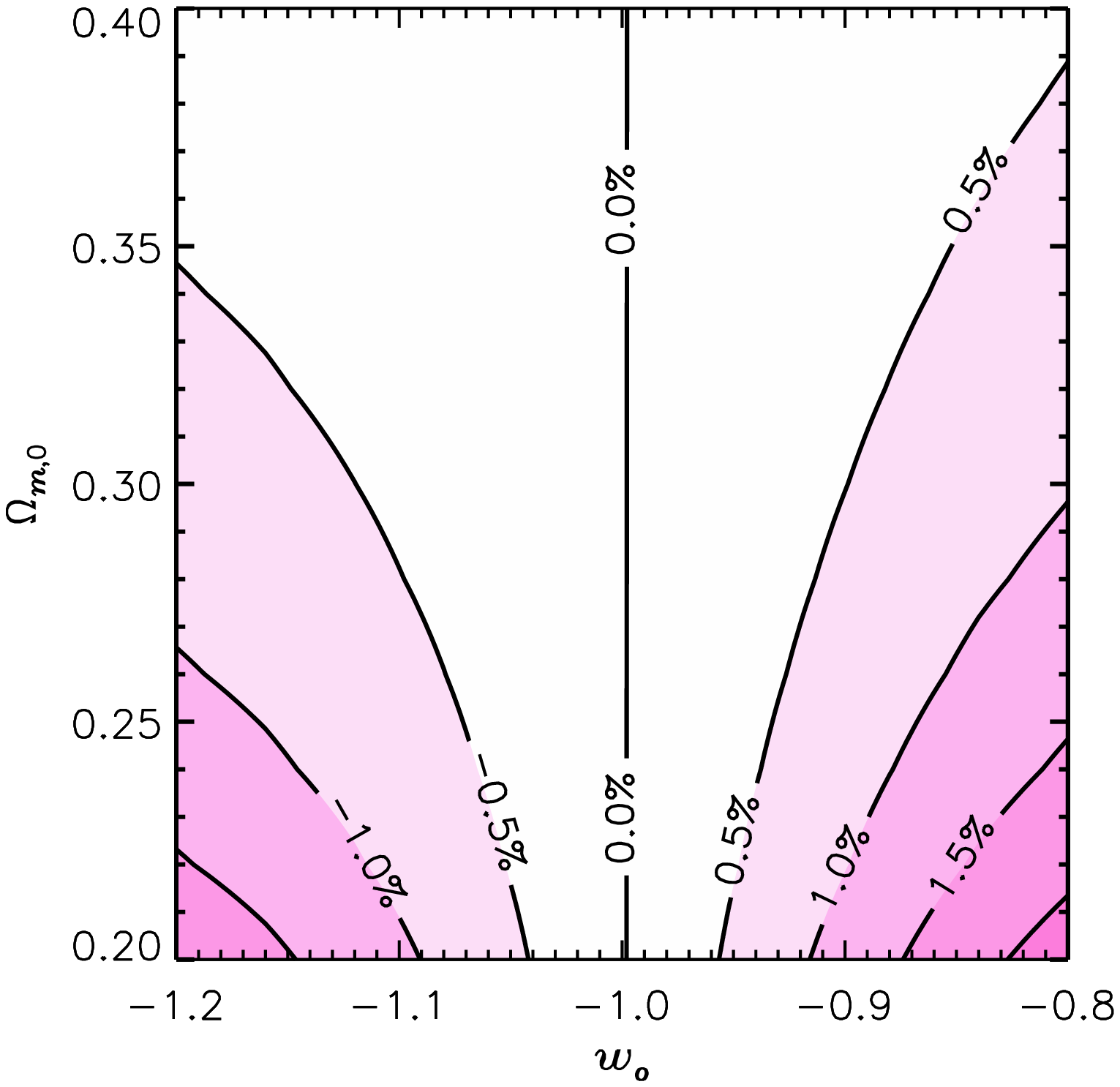}
\caption{Relative error, as a function of the parameters ($\wo, \Omega_{\rm m,0}$), in approximating the growth index of a clustering quintessence model  \cite{SefVer11}
 with Eq. (53) of  \cite{SefVer11} ({\it left panel}) and with Eq. (\ref{eq:gamma_Taylor})  truncated at first order ({\it right panel}). 
The relative error, calculated at $z=0.7$, represents the maximal imprecision in the window surveyed by a Euclid-like survey ($0.7<z<2$).}
\label{fig:sefuverni}
\end{figure}

\subsection{Clustering Dark Energy}\label{sec:clustering}
Dark energy affects the process of structure formation   not only through its equation of state, but also through its
speed of sound. Indeed, if the speed of sound  with which dark energy perturbations propagate drops below  the speed of light, then dark energy
inhomogeneities increase and couple gravitationally to matter perturbations. As a consequence,  they may be detectable on correspondingly better
observable (though typically still large) scales. The influence of  an unstable dark energy component on the clustering of matter can be effectively described  
by switching on the additional degrees of freedom  $\mu$ and $\nu$  in Eq. (\ref{eq:matter_density_fluctuations}).

As an archetype of this class of phenomena, we   consider the clustering dark energy model with vanishing speed of sound introduced in \cite{SefVer11}. 
In this model,  the dark energy {\it EoS}  is assumed to be constant, the speed of sound is effectively null, and we have
\begin{align}
\mu(\x ) = 1 + (1+w) \frac{1-e^{\x}}{e^{\x}} \;, \quad 
\nu(\x ) = - \frac{3(1+w)w(1-e^{\x})}{1+w(1-e^{\x})}
\label{eq:Clustering_mu_nu}
\end{align}
As a consequence, the structural parameters of our formalism are 
\begin{subequations}\label{eq:CQ_M_N}
\begin{align}
\mathcal{M}_0 =  &\; 1\,,\; & \mathcal{M}_n = &\; (-1)^n(1+w) \,, \; n \geq 1
\label{eq:CQ_M} \\
\mathcal{N}_0  =  &\; 1\,,\; & \mathcal{N}_1 = &\; -3 w(1+w)\,,\quad \mathcal{N}_2 = (1+2w)\mathcal{N}_1\,,\quad \mathcal{N}_3=(1+6(1+w)w)\mathcal{N}_1\,,\,\cdots  
\label{eq:CQ_N}
\end{align}
\end{subequations}
while $\mathcal{X}_n$ and $\mathcal{H}_n$ are the same as in Eq.~\eqref{eq:coeff_GR_1}. Using Eq.~\eqref{eq:gamma_n}, it follows immediately that the growth index coefficients are 
\begin{subequations}\label{eq:claqui}
\begin{align}
\gamma_0 &= \tfrac{6\:\! w^2}{5-6\:\! w} \\
\gamma_1 &= \tfrac{3\:\! w^2(75-70\:\! w- 78\:\! w^2 + 72\:\! w^3)}{(5-12\:\! w)(5-6\:\! w)^2}\\
\gamma_2 &= \tfrac{2\:\! w^2(4\:\! 375 - 6\:\! 750\:\! w -13\:\! 800\:\! w^2+33\:\! 480\:\! w^3-5\:\! 544\:\! w^4 + 26\:\! 352 \!\: w^5 + 15\:\! 552\:\! w^6)}{(5-18\:\! w)(5-12\:\! w)(5-6\:\! w)^3}
\end{align}
\end{subequations}
Note that, as for smooth quintessence, the $\Lambda$CDM model is also included in clustering quintessence as the limiting case $w=-1$ (as a matter of fact, by setting $w=-1$ in Eqs.~\eqref{eq:claqui} we finde the coefficients \eqref{eq:gamma012LambdaCDM}).

In Figure~\ref{fig:sefuverni} we compare the exact numerical calculation of the growth index $\gamma(z)$ and  the approximation  (\ref{eq:gamma_Taylor}) 
with coefficients $\gamma_0$ and $\gamma_1$ computed above. 
The relative discrepancy, for the characteristic values of $\Omega_{\rm m,0}$ and $w$ is shown in Figure \ref{fig:sefuverni}. By comparing our parameterization of the growth index with that of Sefusatti and Vernizzi (2011) (left panel of Figure  \ref{fig:sefuverni}), we can appreciate the gain in precision, which is roughly a factor of 2.

\section{Modified Gravity}
\label{sec:modified}
In this section we show how to predict the amplitude  of the growth index   in scenarios where the Einstein fields equations are modified.  
We will first discuss what has become a reference  benchmark  for alternative gravitational scenarios, the DGP model \cite{DGP}, and we will then 
generalize the discussion, by applying the formalism to  Post Parametrized Friedmann models \cite{FerSko10}.

\begin{figure}[t]
\centering
\includegraphics[scale=0.6]{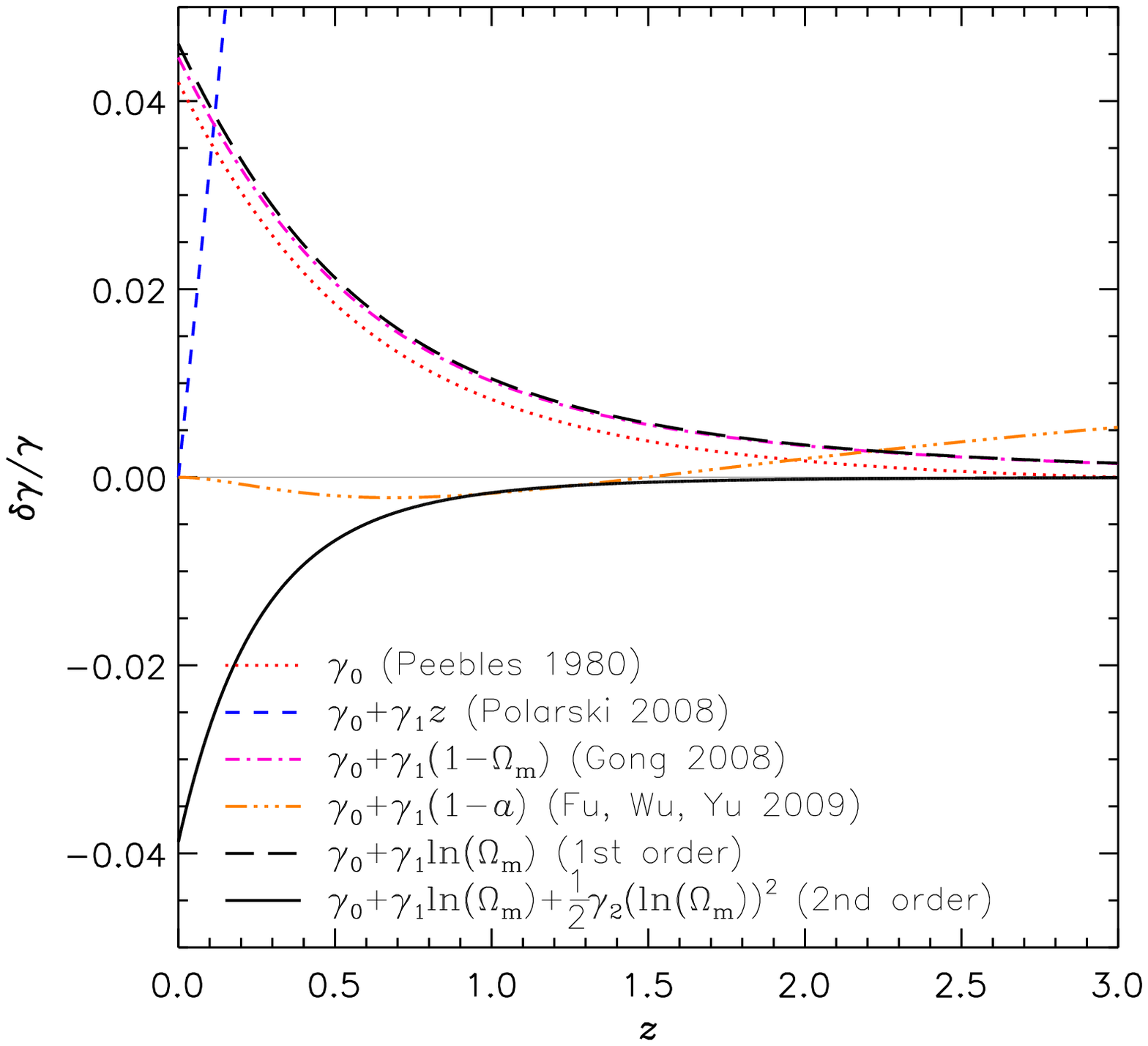}
\caption{The precision of various parameterizations \cite{pee80,pg,Gong,FuWuYu2009} of the growth index is shown for  the flat DGP model with $\Omo=0.213$. 
The precision is estimated as the relative difference with respect  to  the numerical reconstruction of the growth index $\gamma(z)$ in the reference model.}
\label{fig:comparison_DGP}
\end{figure}

\subsection{The growth index as a diagnostic for the DGP model} 
\label{sec:DGP}
We consider the Dvali-Gabadadze-Porrati brainworld model\cite{DGP} in which gravity is modified at large distances due to an extra spatial dimension. 
The scale factor of the  resulting cosmological model evolves according to the following 
modified  Friedmann equation \cite{Deffayet:2001pu}
\begin{align}
H(a)^2 + \frac{k}{a^2}-\frac{1}{r_{c}}\sqrt{H(a)^2+\frac{k}{a^2}} = \frac{8\pi G}{3} \rho_{\rm m}(a)
\label{eq:DGP_Friedmann}
\end{align}
where $r_{c}$ is the length scale at which gravity starts to leak out into the bulk. Neglecting effects of spatial curvature ($k=0$) and defining $\Omega_{r_c}=1/(2 r_c H_0)^2$, the Hubble rate can be expressed as
\begin{align}
H(a) = H_0 \Big[\sqrt{\Omo a^{-3} + \Omega_{r_c}} + \sqrt{\Omega_{r_c}} \Big]
\label{eq:DGP_Hubble}
\end{align}
which implies the constraint  $\Omega_{r_c}=(1-\Omo)^2/4$. By deriving Eq.~(\ref{eq:DGP_Friedmann}) with respect to $\ln a$ and by using the standard energy conservation equation $\dot{\rho}_{\rm m} + 3 H \rhom = 0$, we find, in terms of the fundamental variable of our formalism ($\x$) 
\begin{align}
\x '(\x) =  -\frac{3\:\! (1-e^{\x})}{1+e^{\x}}  \qquad \text{and} \qquad \frac{H'}{H}(\x) = -\frac{3\:\! e^{\x}}{1+e^{\x}}.
\label{eq:DGP_x_h}
\end{align}
from which it follows that the flat DGP model is formally  equivalent to a DE model with  an \eos  varying as $w(\x) =-(1+e^{\x})^{-1}$. The effective Newton constant in the Poisson equation is \cite{Lue06,cdp1,nep}
\begin{align}
G_{\rm eff}(a) = G \Big[1+\frac{1}{3}\Big(1-2 r_c H(a) \big(1+\frac{1}{3}\frac{H'}{H}\big)\Big)^{-1}\Big]\,,
\end{align}
and, after some algebra,  the source term $\mu = G_{\rm eff}/G$ in  Eq.~(\ref{eq:matter_density_fluctuations}) can be expressed as 
\begin{align}
\mu(\x) = 1- \frac{1-e^{2\x}}{3(1+ e^{2\x})} \qquad \qquad \nu(\x) = 1\,.
\label{eq:DGP_mu_nu}
\end{align}
From Eqs. \eqref{eq:DGP_x_h}, \eqref{eq:DGP_mu_nu} and \eqref{eq:struct} we find the following structural parameters
\begin{subequations}\label{eq:DGP_X_H_M_N}
\begin{align}
\mathcal{X}_n &= \lbrace 0,\tfrac{3}{2},0,-\tfrac{3}{4},\cdots \rbrace \;, \quad \mathcal{H}_n = \lbrace -\tfrac{3}{2},-\tfrac{3}{4},0,\tfrac{3}{8}, \cdots \rbrace \\
\mathcal{M}_n &= \lbrace 1,\tfrac{1}{3},0,-\tfrac{2}{3}, \cdots \rbrace \;, \quad \mathcal{N}_n = \lbrace 1,0,0,0,\cdots \rbrace 
\end{align}
\end{subequations}
and, finally, the growth index coefficients
\begin{align}
\gamma_0^{\rm DGP} = \frac{11}{16} = 0.6875 \;, \qquad \gamma_1^{\rm DGP} = -\frac{7}{5\,632} \approx -0.0012 \;, \qquad \gamma_2^{\rm DGP} = -\frac{1051}{22\,528} \approx -0.0467
 \label{gamdgp}
\end{align}

In what follows  we consider the flat DGP model with $\Om=0.213$. With this choice of the density parameter, the DGP model best fits the expansion rate of the {\it reference} $\Lambda$CDM model in the
range $0<z<2$. In this non-standard gravity scenario, the maximal relative error,  when  comparing the parametric reconstruction of the growth rate with numerical results in the redshift range covered by Euclid-like survey,  is  $\delta\gamma/\gamma < 1.5\%$ at order 1, and $\delta\gamma/\gamma < 0.5\%$ at order 2, see Fig \ref{fig:comparison_DGP}.  
To  sub percentage precision we need to expand the growth index  one order further. 
The somewhat degraded precision for the DGP model is   due to the fact that the  Taylor parameterization proposed in this work (cf. the approximation \eqref{eq:gamma_Taylor})  is specifically tailored
for models in which   the growth index minimally deviates  from the $\Lambda$CDM prediction.  It is however impressive how the DGP model, despite its extreme nature,
can still be satisfactorily described by our formalism.

\subsection{The growth index in general models of modified gravity}
\label{sec:gravity}
We will now assume that,  in viable theories of modified gravity,   both background and perturbed observables  have values close  to that  of  the standard 
cosmological model.   That is,  deviations from   $\Lambda$CDM are hypothesized to be small, comparable with current   observational  uncertainties. 
We further assume that  Eq. (\ref{eq:matter_density_fluctuations}) is the master equation  for studying matter collapse on sub-horizon size. In other terms, linear  density fluctuations 
evolve in the quasi-static Newtonian regime. This being the case, 
the structure coefficients of the growth index formalism are given by Eq. (\ref{eq:struct})
and, by a straightforward implementation of  formula (\ref{eq:gamma_n}),  we obtain 
\begin{subequations}\label{eq:MG_gamma_n}
\begin{eqnarray}
\gamma_0      & = & \Bigg[\frac{3(\mu+\frac{d\mu}{d\x }-w)-2\frac{d\nu}{d\x } }{2+3\mu - 6w} \Bigg]_{\x =0} \\
\gamma_{1}   & = & \Bigg[ \frac{-(2-3\mu)\gamma_0^2 -6(\mu + \frac{d\mu}{d\x }-w-2 \frac{dw}{d\x })\gamma_0  - 3\mu + 6\frac{d\mu}{d\x } + 3\frac{d^2\mu}{d\x ^2}-3w-6\frac{dw}{d\x }- 2\frac{d^2\nu}{d\x ^2} }{2(2+3\mu - 12w)}      \Bigg]_{\x =0}  \\
\gamma_{2}   & = & \Bigg[\frac{-(2+3\mu)\gamma_0^3 + 9(\mu+\frac{d\mu}{d\x })\gamma_0^2 - 6(2-3\mu)\gamma_0\gamma_1 + 3\big(2w+6\frac{dw}{d\x }+6\frac{d^2w}{d\x ^2}-3\mu-6\frac{d\mu}{d\x }-3\frac{d^2\mu}{d\x ^2}\big)\gamma_0}{3(2+3\mu -18w)}  \nonumber \\
             &  &  \;\; + \frac{18\big(2w+4\frac{dw}{d\x }-\mu -\frac{d\mu}{d\x } \big)\gamma_1 +3\big(1+w+3\frac{dw}{d\x }+3\frac{d^2w}{d\x ^2}+3\frac{d\mu}{d\x }+3\frac{d^2\mu}{d\x ^2}+\frac{d^3\mu}{d\x ^3}\big)-2\frac{d^3\nu}{d\x ^3}}{3(2+3\mu - 18w)} \Bigg]_{\x =0}
\end{eqnarray} 
\end{subequations}

One can parameterize a large class of modified gravity scenarios by expanding the model dependent quantities $\mu$ and $\nu$ in power series.
Since we are interested in alternative gravitational scenarios that might explain away the dark energy phenomenon, we expect deviations from Einstein's gravity to
become appreciable as $\Omega_{\rm DE}$ starts to diverge from 0.  By assuming both $\mu$ and $\nu$ analytic at $z=\infty$, we can thus expand (see also \cite{FerSko10})
\begin{subequations}\label{eq:taylor_wmunu}
\begin{eqnarray}
w & = & w_i + w_1 \Omega_{\rm DE} + w_2 \Omega_{\rm DE}^2 + \mathcal{O}(\Omega_{\rm DE}^3)\\
\mu & = & 1+\mu_1 \Omega_{\rm DE}+\mu_2\Omega_{\rm DE}^2+\mu_3 \Omega_{\rm DE}^3+ \mathcal{O}(\Omega_{\rm DE}^4)\\
\nu & = & 1+\nu_1 \Omega_{\rm DE}+\nu_2\Omega_{\rm DE}^2+\nu_3 \Omega_{\rm DE}^3+ \mathcal{O}(\Omega_{\rm DE}^4)
\end{eqnarray}
\end{subequations}
where we denoted by $w_i$ the value of the $EoS$ in the epoch of matter domination as in \S \ref{sec:w(a)}. The corresponding growth index coefficients are
\begin{subequations}\label{eq:PMG_gamma_n}
\begin{eqnarray}
\gamma_0 & = & \tfrac{3(1-w_i-\mu_1)+2\nu_1}{5-6w_i} \\
\gamma_1 & = & \tfrac{\gamma_0^2 - 6(1-w_i+2w_1-\mu_1)\,\gamma_0 + 3(1-w_i+2w_1-3\mu_1 + 2\mu_2)+2(\nu_1 - 2\nu_2)}{10-24w_i} \\
\gamma_2 & = & -\tfrac{5\gamma_0^3 -9(1-\mu_1)\gamma_0^2 -6\gamma_0\gamma_1 + 3(3 - 2w_i + 12w_1 - 12w_2-9\mu_1 + 6 \mu_2)\gamma_0 +18(1-2w_i+4w_1-\mu_1)\gamma_1 }{15-54w_i}  \nonumber \\
         &  & +\tfrac{3(1-w_i+6w_1-6w_2-7\mu_1 +12\mu_2 - 6\mu_3) + 2(\nu_1 - 6\nu_2 + 6\nu_3)}{15-54w_i}.
\end{eqnarray}
\end{subequations}

\begin{figure}[t]
\centering
\includegraphics[scale=0.47]{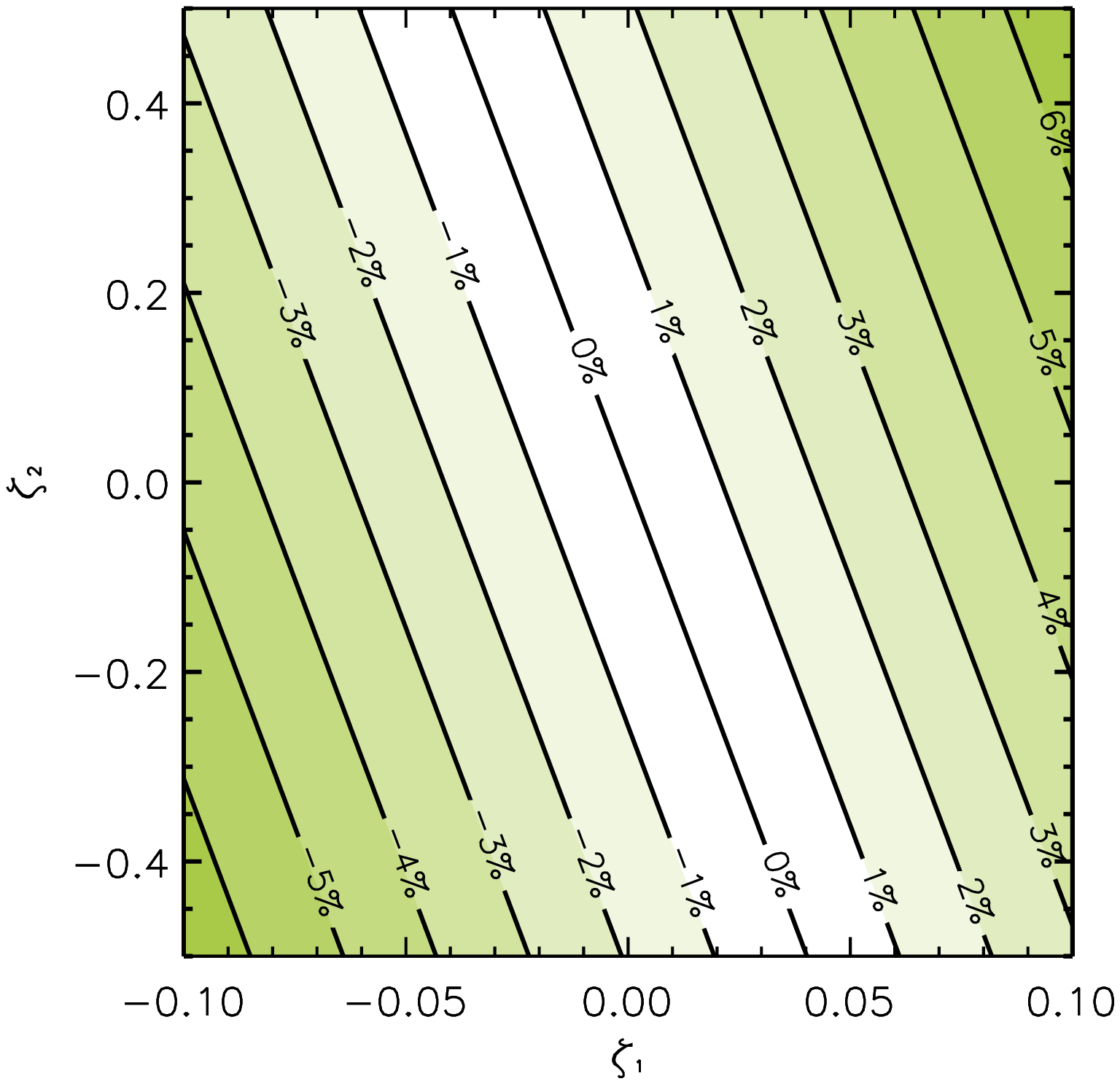}
\includegraphics[scale=0.47]{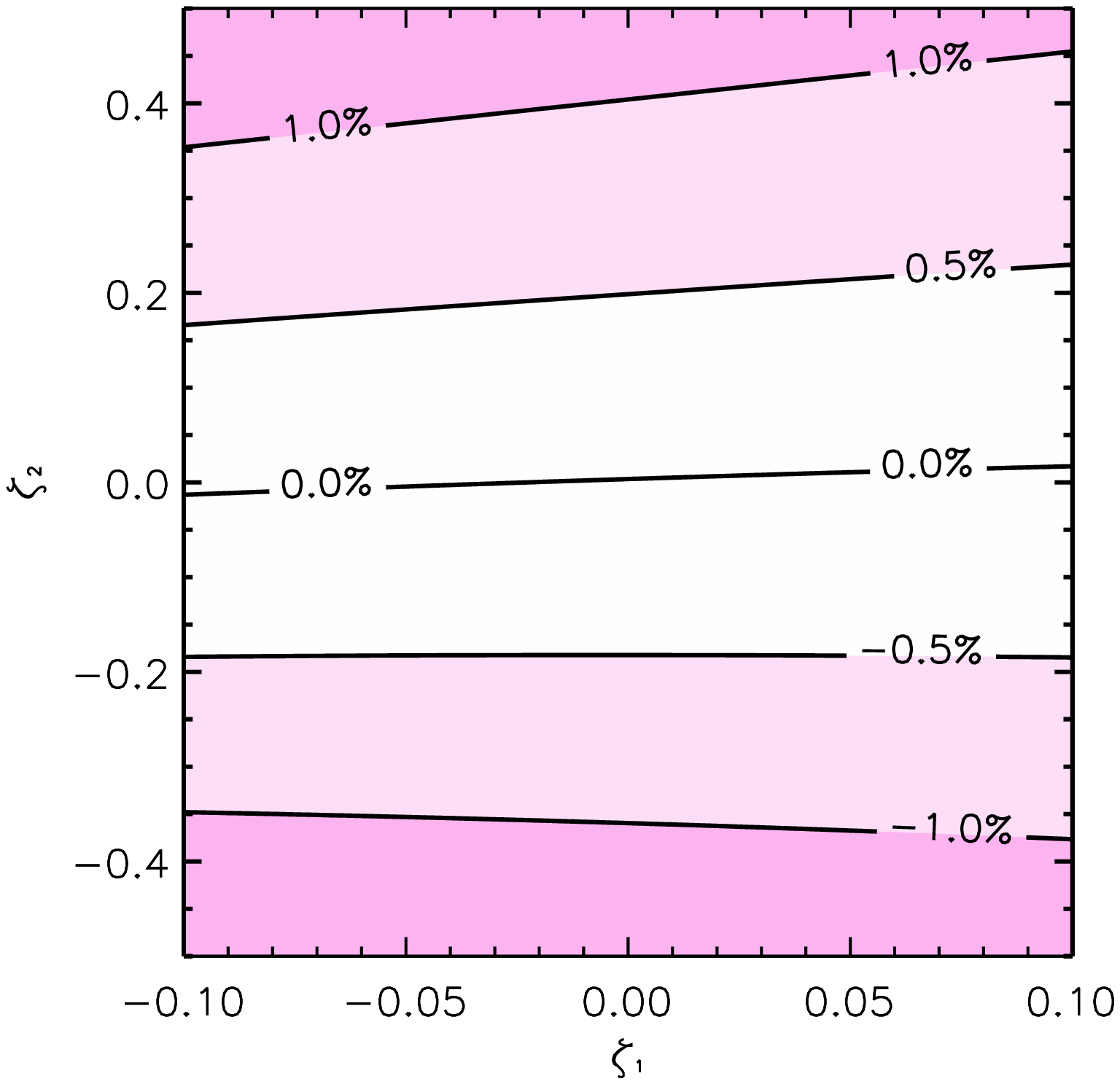}
\caption{{\it Left panel:} relative difference,  at redshift $z=1.5$, between the growth index predicted by the  $\zeta$CDM  model of \cite{FerSko10} and the  $\Lambda$CDM value, for various values of the parameters  $\zeta_1$, and $\zeta_2$. {\it Right panel:} we plot the maximal relative error,   in the redshift interval surveyed by Euclid,  of the first order approximation $\gamma=\gamma_0 + \gamma_1 \ln \Omega_{\rm m}$   
as a function of the amplitude of the parameters  $\zeta_1$ and $\zeta_2$  specifying the $\zeta$CDM  model.     The imprecision is maximal at $z=0.7$,  and it is  smaller than $1\%$  already when  the parameterization  given in 
Eq. \ref{eq:gamma_Taylor}  is truncated  at first order.}
\label{fig:PPF_variation}
\end{figure}

\subsubsection{The $\zeta$CDM framework}
\label{sec:skofe}
As an example of the application of the formalism, we consider the  $\zeta$CDM scenario of \cite{FerSko10}. This is  a one-parameter family of 
models  representing  a large class of modified gravity  theories for which  the fundamental geometric degree of freedom is the metric, 
 the field equations are at most 2nd order in the metric variables and gauge-invariant \cite{sko}. 
 The interesting facet of these non-standard gravity models is that deviations from GR are parameterized by an observable,  the {\it gravitational slip} parameter
\begin{align}
\zeta \equiv 1 - \Phi /\Psi
\label{eq:gravitational_slip}
\end{align}
where $\Phi$ and $\Psi$ are  the  Newton and curvature potentials respectively, both assumed to be  time and scale dependent functions.

In this  non-standard gravity formalism, the background evolution is degenerate with that of the  $w$CDM model of cosmology, that is it   
is effectively described by the Friedmann equation (\eqref{eq:Friedmann}) augmented by a Dark Energy component $\Ol$ with  equation of state parameter $w$, satisfying the usual conservation equation \eqref{eq:conservation}.  This fixes the amplitude of the structural parameters of our formalism to those of \eqref{eq:gamma_coeff_x_h}.

Deviations from General Relativity are  expected only in the perturbed sector of the theory. At first order,
 indeed, this class of models  predicts the following modification of the Newton constant

\begin{align}
G_{\rm eff}(a) = G \big(1- \zeta(a)\big),
\end{align}

where it is assumed that on small cosmic scales, well inside the horizon, any scale dependence in the gravitational slip parameter can be neglected. 
This in turns implies that the growth of matter perturbations, in these scenarios, can be described by inserting 
$\mu = 1-\zeta$ and $\nu = 1$ into Eq. \ref{eq:f_H_General}. Since,  at early epochs deviations from General Relativity are  constrained by Big Bang Nucleosynthesis measurements, it is a fair 
hypothesis to assume  that $\zeta(t)$ is a smooth function which  deviates  from zero as soon as the dark energy comes to dominate the overall energy budget of the universe. As a result, the  gravitational slip parameter can be expanded as \cite{FerSko10}%
\begin{equation}
\zeta = \zeta_1 \Omega_{\rm DE}+\zeta_2 \Omega_{\rm DE}^2 +\zeta_3 \Omega_{\rm DE}^3  + .....
\label{eq:zeta}
\end{equation}
and the overall effects of non-standard gravity are governed by the set of discrete parameters $\zeta_i$. 
From Eqs.~\eqref{eq:taylor_wmunu} and \eqref{eq:PMG_gamma_n} it is now easy to read off the value of the growth indices in the $\zeta$CDM model:
\begin{subequations}\label{eq:PPF_gamma_n}
\begin{eqnarray}
\gamma_0      & = & \tfrac{3(1-w_i+\zeta_1)}{5-6w_i} \\
\gamma_{1}   & = & \tfrac{\gamma_0^2 - 6(1-w_i+2w_1+\zeta_1 )\gamma_0 +3(1-w_i+2w_1+3\zeta_1 -2\zeta_2))}{10-24w_i} \\
\gamma_{2}   & = &   -\tfrac{5\gamma_0^3 -9(1+\zeta_1)\gamma_0^2 -6\gamma_0\gamma_1  + 3(3 -2w_i +12w_1-12w_2 +9\zeta_1 -6\zeta_2)\gamma_0 + 18(1-2w_i+4w_1+\zeta_1)\gamma_1}{15-54w_i}  \nonumber \\
           & &  +\tfrac{3(1-w_i+6w_1-6w_2+7\zeta_1 - 12\zeta_2 + 6\zeta_3)}{15-54w_i}
\end{eqnarray} 
\end{subequations}

The amplitude of the non-standard signals expected in this alternative gravitational  scenario  is shown in the left panel of Figure \ref{fig:PPF_variation} where we display 
the distortions which a possibly  non-null value of the $\zeta$CDM parameters $\zeta_1$ and $\zeta_2$  induce  on  the growth index. 
Also the accuracy with which the growth index is reconstructed by our formalism is shown (left panel). This last plot confirms that systematic uncertainties are below the 
threshold of the Planck statistical errors  over a region of the parameter space  ($\zeta_1$, $\zeta_2$) which is sufficiently large to be physically interesting.

\begin{table}[h]
\centering
\begin{tabular}{|l|l|c|c|}
\hline
Label      & Reference                                        & $z$  & $f\sigma_8$    \\
\hline
\hline
THF        & Turnbull {\it et~al.} (2012) \cite{TurHudFel12}  & 0.02 & $0.40\pm 0.07$  \\
\hline
DNM        & Davis {\it et~al.} (2011) \cite{DavNusMas11}     & 0.02 & $0.31\pm 0.05$  \\
\hline
6dFGS      & Beutler {\it et~al.} (2012) \cite{BeuBlaCol12}   & 0.07 & $0.42\pm 0.06$  \\
\hline
2dFGRS     & Percival {\it et~al.} (2004) \cite{PerBurHea04}, Song \& Percival (2009) \cite{SonPer09}   & 0.17 & $0.51\pm 0.06$  \\
\hline
2SLAQ      & Ross {\it et~al.} (2007) \cite{RosAngSha07}    & 0.55 & $0.45\pm 0.05$  \\
\hline
SDSS       & Cabr\'{e} {\it et~al.} (2009) \cite{CabGaz09}    & 0.34 & $0.53\pm 0.07$  \\
\hline
SDSS II    & Samushia {\it et~al.} (2012) \cite{SamPerRac12}  & 0.25 & $0.35\pm 0.06$  \\
           &                                                  & 0.37 & $0.46\pm 0.04$  \\
\hline
BOSS       & Reid {\it et al.} (2012) \cite{ReiSamWhi12}      & 0.57 & $0.43\pm 0.07$  \\
\hline
WiggleZ    & Contreras {\it et~al.} (2013) \cite{ConBlaPoo13} & 0.20 & $0.40\pm 0.13$ \\
           &                                                  & 0.40 & $0.39\pm 0.08$  \\
           &                                                 &  0.60 & $0.40\pm 0.07$  \\ 
           &                                                 &  0.76 & $0.48\pm 0.09$  \\ 
\hline
VVDS       & Guzzo {\it et al.} (2008) \cite{GuzPieMen08}, Song \& Percival (2009) \cite{SonPer09} & 0.77 & $0.49\pm0.18$  \\ 
\hline
VIPERS     & De la Torre {\it et al.} (2013) \cite{TorGuzPea13}& 0.80 & $0.47\pm0.08$ \\
\hline
\end{tabular}
\caption{Compilation of currently available growth rate data.}
\label{tab:data}
\end{table}

\section{Data analysis in the growth index  parameter space}
\label{sec:constraining}
Besides being instrumental in increasing the accuracy with which the growth rate is reconstructed from data, the $\gamma$-formalism  introduced in the previous sections 
also serves as a guide in interpreting empirical results directly in terms of dark energy models. This is shown in this section, 
where we  confront the growth index predictions with current data and data simulations for a Euclid-like survey. After describing  our data analysis strategy,  we  show here how  we  test whether the $\Lambda$CDM model correctly describes available data about the linear growth rate of structures and how we 
use the growth index parameter space  $\gamma_0-\gamma_1$ to analyze and draw statistical conclusions on various  non-standard gravity scenarios, in a manner which is independent from the specific details of the expansion history of the universe.

\subsection{Testing the $\Lambda$CDM predictions in the perturbed sector}
\label{sec:null_hypothesis}

We first focus on the  constraints that present day observations  set on the growth indices $\gamma_i$. These are derived by  computing the likelihood $\mathcal{L}$ of the data shown in Table \ref{tab:data}  given the  model  in  Eqs. \eqref{eq:f_approximation} and  \eqref{eq:gamma_Taylor}. To this purpose we minimize the $\chi^2=-2\ln(\mathcal{L})$  function 
\begin{align}
\chi(\boldsymbol{\gamma},\mathbf{p})^2= \sum_{i=1}^{N} \Bigg( \frac{\big(f\sigma_8\big)_{\! \rm obs}(z_i) - f(\boldg,\mathbf{p},z_i)\sigma_8(\boldg,\mathbf{p},z_i)}{\sigma_i} \Bigg)^2
\end{align}
where $\sigma_i$ is the uncertainty in the growth data, $\mathbf{p}=(\sigma_{8,0},\Omega_{\rm m,0},\wo,\wa)$ is the set of parameters that, except for $\sigma_{8,0}$, fix the background expansion, $\boldg =(\gamma_0,\gamma_1,...)$ are the growth indices introduced in Eq.~\eqref{eq:gamma_Taylor} and
\begin{align}
f(\boldg,\mathbf{p},z) &=  \Omega_{\rm m}(\mathbf{p},z)^{\sum_i \gamma_i \big(\ln \Om(\mathbf{p},z)\big)^i/i!} \label{eq:deff0} \\
\sigma_8(\boldg,\mathbf{p},z) &= \sigma_{8,0} D(\boldg,\mathbf{p},z) = \sigma_{8,0}\;\! e^{\int_0^z \tfrac{f(\boldg,\mathbf{p},z')}{1+z'} dz'} .
\label{eq:deff}
\end{align}
where $D=\delta(t)/\delta(t_0)$ is the growth factor. In this paper we restrict our analysis to the first two growth indices, i.e.~we set $\boldg=(\gamma_0,\gamma_1)$. We also assume the $f\sigma_8$ measurements in Table \ref{tab:data} to be independent. A more sophisticated error analysis might eventually results  in minor  changes in our quantitative conclusions, but will have no impact on their 
physical interpretation.   It is well known that $f\sigma_8$   cannot be estimated from data without picking a 
particular model, or at least a parametrization, for gravity, i.e., for the quantity being tested  \cite{motta}. Despite the strong prior on the underlaying gravitational model, there is however evidence that 
the estimated values of this observable  depend  negligibly on the distance-redshift conversion model, that is for sensible variations
of the background  parameter $\Omo$ in a flat $\Lambda$CDM model, the variation of the estimation is  well below the statistical error  \cite{TorGuzPea13,ConBlaPoo13}. 
This should guarantee  that data can be meaningfully compared against models with distinct  background evolution  from that assumed to estimate the observable. 

\begin{figure}[t]
\centering
\includegraphics[scale=0.5]{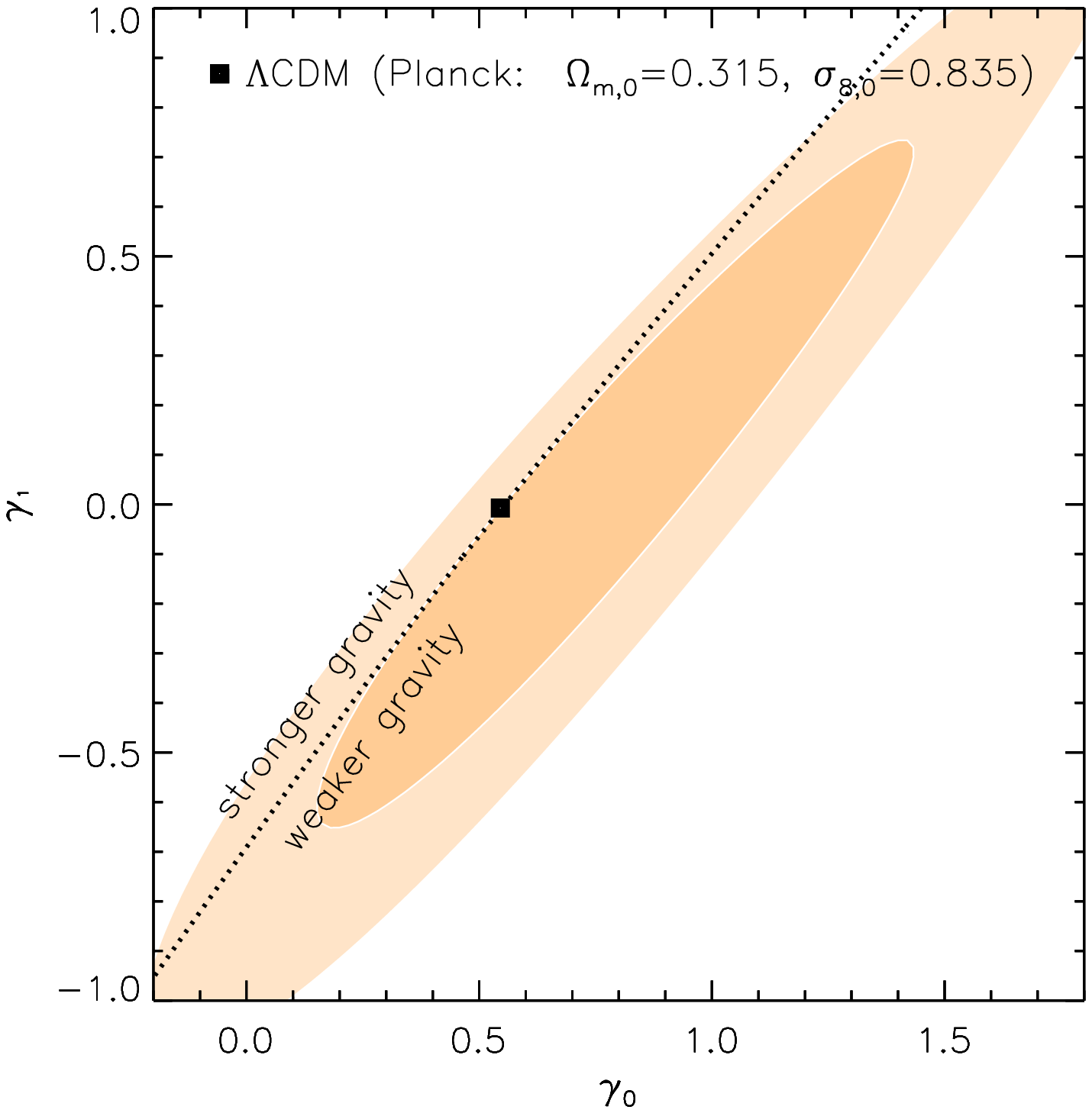}
\includegraphics[scale=0.5]{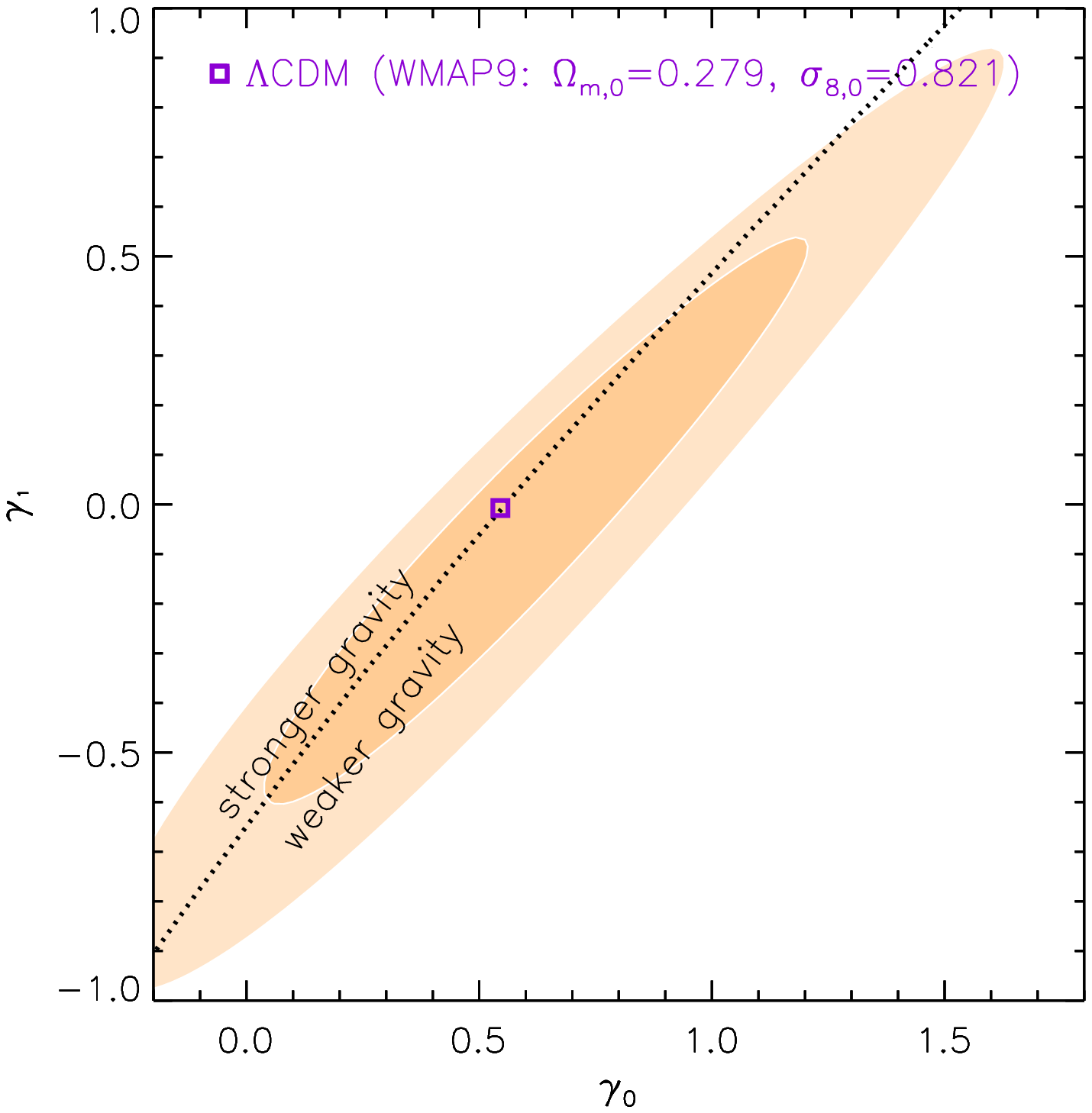}
\caption{{\it Left panel:} $68\%$ and $95\%$ confidence levels in the growth index parameter space $\gamma_0- \gamma_1$ (orange filled regions) obtained from  the compilation of data shown in Table \ref{tab:data} and by  using,  as fiducial for the evolution of the background metric,  the expansion rate of the {\it reference} $\Lambda$CDM model ($\Omo=0.315$, $\sigma_{8,0}=0.835$). The marginalized best-fitting values are $\gamma_0=0.74^{+0.44}_{-0.41}$ and $\gamma_1=0.01^{+0.46}_{-0.46}$. The growth indices theoretically expected in the fiducial $\Lambda$CDM model ($\gamma_0=0.55$, $\gamma_1=-0.007$) are shown by the filled black square. The black dotted line, defined by Eq.~\eqref{eq:stronger_weaker}, shows the partition of the $\gamma_0 - \gamma_1$ plane into regions where growth is amplified/suppressed with respect to the fiducial model.  {\it Right panel:} The same but using as fiducial the $\Lambda$CDM model with WMAP9 values ($\Omo=0.279$, $\sigma_{8,0}=0.821$). The $\Lambda$CDM model is represented by the empty purple square.}
\label{fig:planck_wmap9}
\end{figure}

The likelihood analysis is carried out by choosing a prior model (hereafter called {\it fiducial}) for the evolution of $\Om$ and $\sigma_8$ in Eqs. (\ref{eq:deff}).  
To this purpose we choose the   {\it reference} $\Lambda$CDM model i.e. a  flat $\Lambda$CDM model  with $Planck$ parameters  $\mathbf{p}=(0.835,0.315,-1,0)$ \cite{Ade:2013zuv}. The resulting likelihood contours for $\gamma_0$ and $\gamma_1$ are shown in the left panel of Figure \ref{fig:planck_wmap9}. 
By marginalizing in the growth index parameter space 
 we find that, at $68\%$ confidence level (c.l.), the relative precision on the leading and first order growth indices is $\gamma_0=0.74^{+0.44}_{-0.41}$ and $\gamma_1=0.01^{+0.46}_{-0.46}$. 
On the right panel of Figure \ref{fig:planck_wmap9} we show the same analysis assuming WMAP9 background values $\mathbf{p}=(0.821,0.279,-1,0)$. In this case, the 1-dimensional marginalized $68\%$ confidence levels are $\gamma_0=0.58^{+0.40}_{-0.38}$ and $\gamma_1=-0.06^{+0.39}_{-0.38}$. The likelihood contours are appreciably smaller if the likelihood analysis is carried out 
in the WMAP9 fiducial, owing to the fact that the statistical analysis depends on the background model. We will see in the next section how to factor out the specific choice of the background model 
from the interpretation of growth rate data.

The traditional  way of exploiting the growth index  formalism is to use it as a tool for a consistency test of the scenario proposed to model the background kinematics of the universe. The goal is to check whether a given   gravitational  model  that explains the observed cosmic  expansion rate   also predicts the correct growth rate  of linear structures.  For example,  one needs to verify that  the most likely amplitude of the growth index parameters derived from observations of the linear growth of structures is not in conflict with those predicted on the basis of the DE {\it  EoS} parameters  $\wo$ and $\wa$ which best fit expansion rate data.  In  absence of major observational systematics, any possible mismatch between measured and expected values of the growth index, would be the smoking gun of new gravitational physics.  In the opposite case,  growth data provide additional constraints on dark energy parameters. 

Figure \ref{fig:planck_wmap9} shows that the growth index predicted on the basis of   $\wo=-1$ and $\wa=0$, i.e.~the {\it EoS} values of the reference $Planck$ $\Lambda$CDM model agrees  with results from the likelihood analysis of linear growth rate data coming from a variety of low redshift surveys of the large scale structure of the universe.   Shouldn't this be the case, 
one could question either the unbiased nature of the data analysis in both sectors, background and perturbed, either the effectiveness of the standard description of gravity in terms of a $\Lambda$CDM framework.

A first immediate advantage of interpreting growth history data in the growth index plane, is that this parameter space  facilitates the  interpretation of  critical information encoded in the likelihood function. 
Specifically, it allows to classify alternative dark energy models (each labeled by the pair of coordinates $(\gamma_0,\gamma_1)$)  as either generating more or less growth of structures with respect to 
the chosen fiducial. The line separating these two characteristic regions is shown in Figure  \ref{fig:planck_wmap9} and it is computed by imposing
\begin{align}
D(\boldg, \bar{\mathbf{p}}, z_{\rm init}) = D(\bar{\boldg},\bar{\mathbf{p}},z_{\rm init})
\label{eq:stronger_weaker}
\end{align}
where $\bar{\boldg}$ are the growth indices of the fiducial model (that is  $\bar{\boldg}=(6/11,-7/2057)$ in Fig \ref{fig:planck_wmap9}), $\bar{\mathbf{p}}$ are the background parameters of the fiducial model and $z_{\rm init}$ is the  initial redshift at which perturbations are conventionally assumed start growing. The region of more growth in the $\gamma$-plane is defined as the region where a density perturbation
(whose amplitude is normalized to unity today)  had, at the initial redshift,  a smaller amplitude than that predicted  in the fiducial model ($D(z_{\rm init})< \bar{D}(z_{\rm init})$)  whereas 
regions of weaker gravity are the $\gamma$ loci   where the amplitude of the  perturbation was larger  ($D(z_{\rm init}) > \bar{D}(z_{\rm init})$). 
Note that the orientation of the  line separating these two regions depends only negligibly on the chosen initial redshift  $z_{\rm init}$ (for the sake of illustration we have set $z_{\rm init} =100$ in Figure \ref{fig:planck_wmap9}). Our analysis shows that  current data mildly favor models in which the strength of gravitational interactions is weaker than what is predicted in the {\it reference} $\Lambda$CDM model.
This result also confirms a conclusion of \cite{samu}. As compared to the way perturbations grow within purely matter-dominated models,  it seems that  data tend to prefer a  suppression mechanism  more efficient than that provided by the cosmological constant.

\begin{figure}[t]
\centering
\includegraphics[scale=0.7]{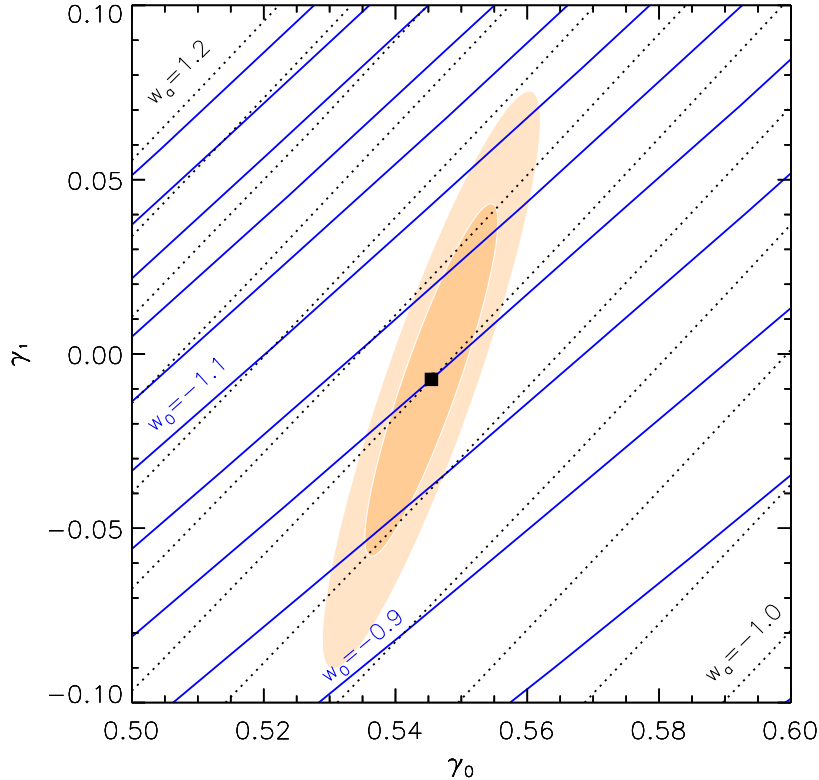}
\caption{Forecasted $68\%$ and  $95\%$ confidence levels in the growth indices parameter space $\gamma_0$ - $\gamma_1$  (orange filled regions) for a Euclid-like survey, assuming as fiducial cosmological model the  {\it reference} $\Lambda$CDM model  ($\Omo=0.315$, $\sigma_{8,0}=0.835$). The black dotted and blue solid lines show predictions of a smooth quintessence model, that is a Dark Energy component with variable {\it EoS} of parameters $\wo$ and $\wa$ (see Eq.~\ref{eq:wi_finite}). These parameters span the range $[-1.3,-0.85]$ and $[-0.6,1.4]$ respectively.  The spacing between adjacent lines is 0.05 and 0.2 respectively. The growth indices theoretically expected in the fiducial $\Lambda$CDM model is shown by a filled black square.
}
\label{fig:euclid_wowa}
\end{figure}

\subsection{Discriminating DE models in the $\gamma_0$ and $\gamma_1$ parameter space}
\label{sec:mapping}

Once a  fiducial  for the background evolution is  chosen to compute the likelihood function, what conclusions can we draw on the viability of  gravitational  models other than the fiducial itself?
We will now see that, since  the dependence of the growth index on  the relevant cosmological  parameters is explicitly taken into account in our analysis scheme, 
the  likelihood in the $\gamma_0 - \gamma_1$ plane, besides allowing to reject the null hypothesis that  the fiducial is compatible with data, also allows  to set constraints  on  alternative 
cosmological scenarios. 
In other terms, it is possible to exploit model dependent likelihood contours in the $\gamma_0-\gamma_1$ plane to tell apart also those theoretical models characterized
by an evolution of the background sector $\mathbf{p}$ which is distinct from that of the fiducial model $\bar{\mathbf{p}}$ itself. 

The growth indices $\gamma_0$ and $\gamma_1$  are  model dependent quantities which can be estimated only once a specific background fiducial model for the evolution of $\Om$  is chosen (cf.~Eq.~\ref{eq:deff0}). As a consequence, the growth indices inferred in distinct  background models cannot be directly compared. For example, consider the likelihood contours in the plane $\gamma_0-\gamma_1$ obtained by assuming the {\it reference} $\Lambda$CDM model as fiducial (Figure \ref{fig:planck_wmap9}). One cannot constrain the DE $EoS$ parameters $\wo$ and $\wa$ by simply computing the amplitude of the coefficients $\gamma_{0}$ and $\gamma_{1}$ expected in an effective DE model with $EoS$ parameters $\wo$ and $\wa$  (cf.~\S \ref{sec:w(a)}) and by confronting these theoretical values with the empirical likelihood. Indeed,  the specific  growth rate history  of a DE model is not entirely
captured by the growth indices, part of the information being locked in the scaling of the background density parameter  $\Om$.

We overcome this critical pitfall  by  computing  $\boldg^*$,  the amplitude of the effective
growth indices in the fiducial background, that is the exponent that once inserted in  (\ref{eq:f_approximation}) together with the  matter density parameter of the fiducial model adopted in the likelihood analysis ($\bar{\Omega}_{\rm m}$), allows to match  the  scaling of the growth rate expected in the specific gravitational model under consideration. This is equivalent to enforcing the following identity
\begin{align}
f(\boldg^*,\bar{\mathbf{p}},z) = f(\boldg, \mathbf{p},z).
\label{eq:f_approximation2}
\end{align}   
By means of this transformation strategy  we  factor out the effect of expansion from the analysis of growth rate histories. 
We will  illustrate this feature by simulating the constraints that the growth index parameters expected from a next generation survey such as the space mission Euclid will put on the background dark energy parameters $\wo$ and $\wa$. 

The Euclid   mission is designed to survey, in  spectroscopic mode, $\sim 5\cdot 10^7$ galaxies  
in  the redshift range $0.5<z<2.1$ and in a sky area of $\sim 15000$ deg$^2$ \cite{euclid2}.
We simulate the expected growth data $\gamma_{\rm obs}$ assuming as fiducial,  the reference  $\Lambda$CDM model. 
To this purpose, we simply split the redshift range $0.7<z<2$ into 14 intervals, and we predict $\gamma_{\rm obs}=\ln f/\ln \Om$ by using Eqs.~\eqref{eq:relations_SDE}, \eqref{eq:relations_SG} and \eqref{eq:f_H_x}. We finally  assume that the relative error on the observable, in each interval, is that corresponding to the Euclid figure of merit  listed in table 2.2 of \cite{euclid2}, i.e.~$\delta\gamma/\gamma=1\%$. 
The resulting likelihood contours  in the   $\gamma_0 - \gamma_1$ plane, obtained via a standard $\chi^2$ analysis, are shown in  Figure~\ref{fig:euclid_wowa}.   
Notice that  the growth index figure of merit, defined as the inverse of the surface  of the $68\%$  likelihood contour in the $\gamma_0-\gamma_1$ plane, is expected to increase by a factor $\sim 550$  when compared to that deduced from current constraints (see Figure \ref{fig:planck_wmap9}).

In Figure~\ref{fig:euclid_wowa} we also show the effective growth indices  ($\gamma_0^*,\gamma_1^*)$ predicted by various DE models (labeled by the $EoS$ parameters $\wo$ and $\wa$) obtained by means of the transformation law  \eqref{eq:f_approximation2}. In practice we compute these effective growth indices ($\gamma_0^*,\gamma_1^*$) by minimizing numerically the integral
\begin{align}
\int \Big(f(\boldg^*,\bar{\mathbf{p}},z) - f(\boldg,\mathbf{p},z)\Big)^2 dz\,,
\end{align}
over the redshift range covered by observational data (i.e.~$[0,0.8]$ for current data and $[0.7,2]$ for Euclid-data simulations). We have verified that this mapping is sufficiently precise over all the redshift intervals where acceleration effects are observable. For $0.7< z<2$ typical errors of order $0.3\%$ arise for the most extreme values of $\wo$ and $\wa$ shown in Figure~\ref{fig:euclid_wowa}. Therefore, this error is largely negligible with respect to the precision of the constraints.
Note that while varying $\wo$ and $\wa$, we have kept fixed $\Omo$, that is we overlook any possible degeneracy between the DE \eos and matter density parameters entering the expansion rate $H(z)$. 
This essentially  because the relative variation induced in the distance modulus $\mu(z)=5\,\text{log}_{10}(d_{L}(z)/\text{Mpc})+25$ by this simplifying assumption is less than $0.2\%$ in the domain of interest.
Overall, Figure~\ref{fig:euclid_wowa} shows how measurements in the perturbed sector help to tighten the constraints on background cosmological parameters, and, ultimately to discriminate among DE models. 
Specifically, we predict  that growth rate data will provide independent constraints on the value of $w_0(/w_a)$ characterized by  a relative(/absolute) precision of  $1\% (/0.5)$.

\begin{figure}[t]
\centering
\includegraphics[scale=0.5]{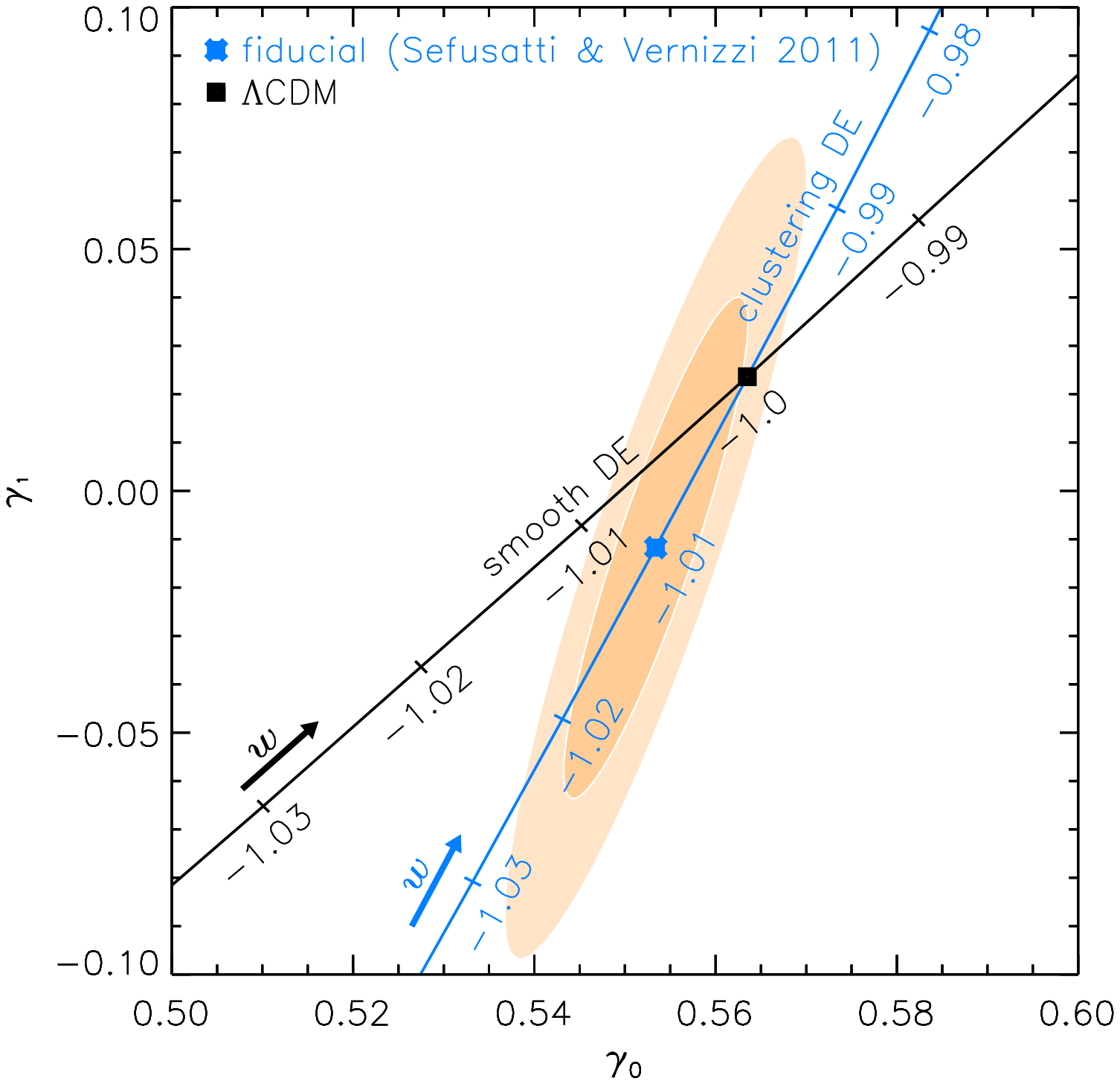}
\includegraphics[scale=0.5]{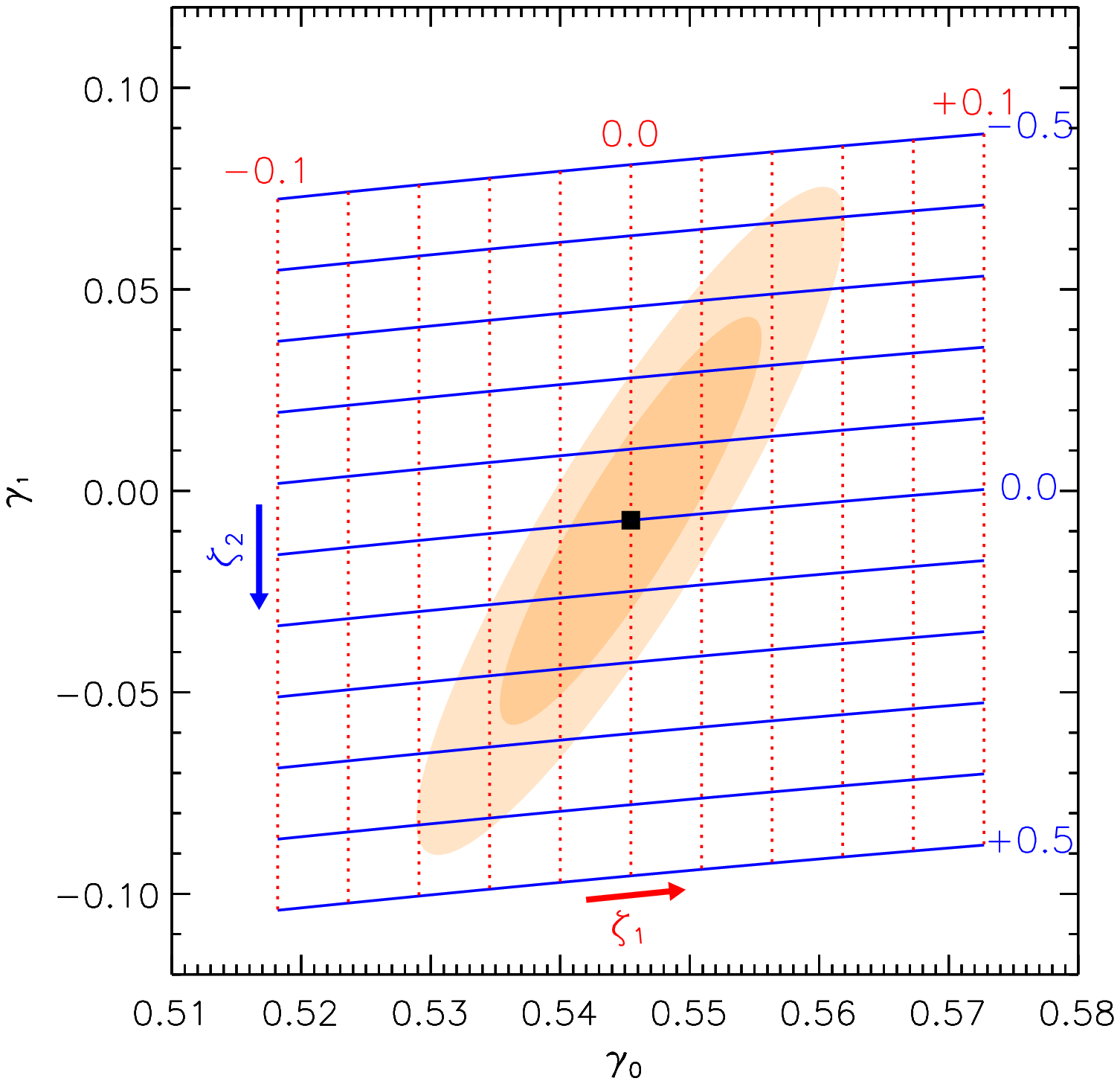} 
\caption{{\it Left panel:} Forecasted $68\%$ and  $95\%$ confidence levels in the growth indices parameter space $\gamma_0$ - $\gamma_1$  (orange filled regions) for a Euclid-like survey, assuming as  fiducial of a flat clustering quintessence model  \cite{SefVer11} with parameters $\Omo=0.315$ and $w=-1.01$. The theoretically predicted values $(\gamma_0,\gamma_1)=(0.5534,-0.0118)$ for this fiducial model are shown by a light blue cross. We also show the effective growth indices (in the chosen fiducial) for  smooth and clustering quintessence models with constant  $EoS$ parameter $-1.03>w>-0.98$ (black  and blue lines respectively). 
The spacing between points is $0.01$  with $w$ values increasing along the direction specified by the arrow.
{\it Right panel:} The same as the left panel but here we have overplotted the amplitude of the growth indices $\gamma_0$ and $\gamma_1$ predicted by the $\zeta$CDM  model of  \cite{FerSko10} as a function $\zeta_1=[-0.1,0.1]$ (red dashed isocontours) and $\zeta_2=[-0.5,0.5]$ (blue full isocontours). The spacing between adjacent lines is $0.02$ and $0.1$ respectively. The fiducial model is the {\rm reference} $\Lambda$CDM model.
}
\label{fig:same_background}
\end{figure}

The remapping method illustrated by Eq.~\eqref{eq:f_approximation2} provides a technique with which to tell apart gravitational models with identical background evolutions.  
In other terms we can use the growth index parameter space  to resolve the background degeneracy of two
dark energy models, i.e.~two models predicting the same background expansions, but different  linear growth histories.
To illustrate this feature, we consider two possible scenarios whose background evolutions are degenerate with that of the $\lambda$CDM model: the 
clustering quintessence and  the $\zeta$CDM  models.

We first consider the case in which future Euclid data are simulated in a clustering  quintessence model,  a cosmological model  in which  matter  is supplemented by a quintessence component 
with null sound speed, i.e.~the model of \cite{SefVer11} discussed  in \S \ref{sec:clustering}. Specifically,  we  forecast  $\gamma_{\rm obs}$ by using Eqs.~\eqref{eq:Clustering_mu_nu}, \eqref{eq:relations_SG} and\eqref{eq:f_H_x} and by further imposing  that the background expansion is effectively described in terms of the constant DE {\it EoS}  parameter  $w=-1.01$.  
The  likelihood contours  in the   $\gamma_0 - \gamma_1$ plane, obtained by adopting as fiducial model the same clustering quintessence model used to simulate data, 
are displayed on the left panel of Figure~\ref{fig:same_background}.   
We now show how to use  these measurements in the $\gamma$ plane to tell apart the clustering quintessence from canonical smooth quintessence.  To this purpose
we calculate the  effective growth indices  $\gamma_0^*$ and $\gamma_1^*$  in two different theoretical models of dark energy, the clustering  quintessence and the smooth quintessence models
(both identified, for simplicity,  via their constant $EoS$ parameter $w$). 
 They are calculated by using  Eq.~\eqref{eq:f_approximation2} to  map $(\gamma_0,\gamma_1)$ of Eqs.~\eqref{eq:gammanw(a)CDM} and \eqref{eq:claqui} respectively into the effective values $(\gamma_0^*,\gamma_1^*)$ for the fiducial background $\bar{\mathbf{p}}=(0.835,0.315,-1.01,0)$ used to analyze growth data.
By comparing likelihood results vs. the  predicted  amplitude of the growth index, we can appreciate how the smooth DE model with the same background expansion as the {\it true} cosmological model
(in this context,  the clustering quintessence model) is clearly ruled out at 95\% c.l.~by growth rate data.  Specifically,  if the effective {\it EoS} deviates from the reference value $w=-1$ by at least $1\%$ ($w<-1.01$ or $w>-0.99$)   a Euclid-like survey has enough resolving power to  discriminate between smooth and clustering quintessence.

As an additional example,  on the right panel of Figure~\ref{fig:same_background} we show  the constraints that an  Euclid-like survey can set  on the slip parameter entering the 
$\zeta$CDM model reviewed  in \S \ref{sec:skofe}. To this purpose we  simulate Euclid observations assuming the  {\it reference} $\Lambda$CDM model and we then reconstruct the likelihood assuming this very same model as fiducial.  We then compare this statistic to prediction from the $\zeta$CDM model, that is to the values  
$\gamma_0$ and $\gamma_1$ of Eqs.~\eqref{eq:PPF_gamma_n} computed for different values of the slip parameters $\zeta_1$ and $\zeta_2$ defined in equation \eqref{eq:zeta} (for simplicity we here 
explore only models for which   $w_i=-1$ and $w_1=w_2=....=0$). Note that,  in this particular case, owing to the fact that  the background evolution does not change from model to model, 
the remapping procedure is superfluous. The right panel of Figure~\ref{fig:same_background}  displays the performances of an Euclid-like survey in detecting  possible deviations of the 
slip parameter $\zeta$  from its GR value, {\it i.e} from zero. We can conclude that data will have enough resolving power  to exclude,  at the 95\% c.l., models predicting 
a parameter $\zeta_1$ larger than 0.025 , or in other terms data can detect a relative deviation between the Newtonian and the curvature potential if it is larger than about 2.5\%.

\begin{figure}[t]
\centering
\includegraphics[scale=0.5]{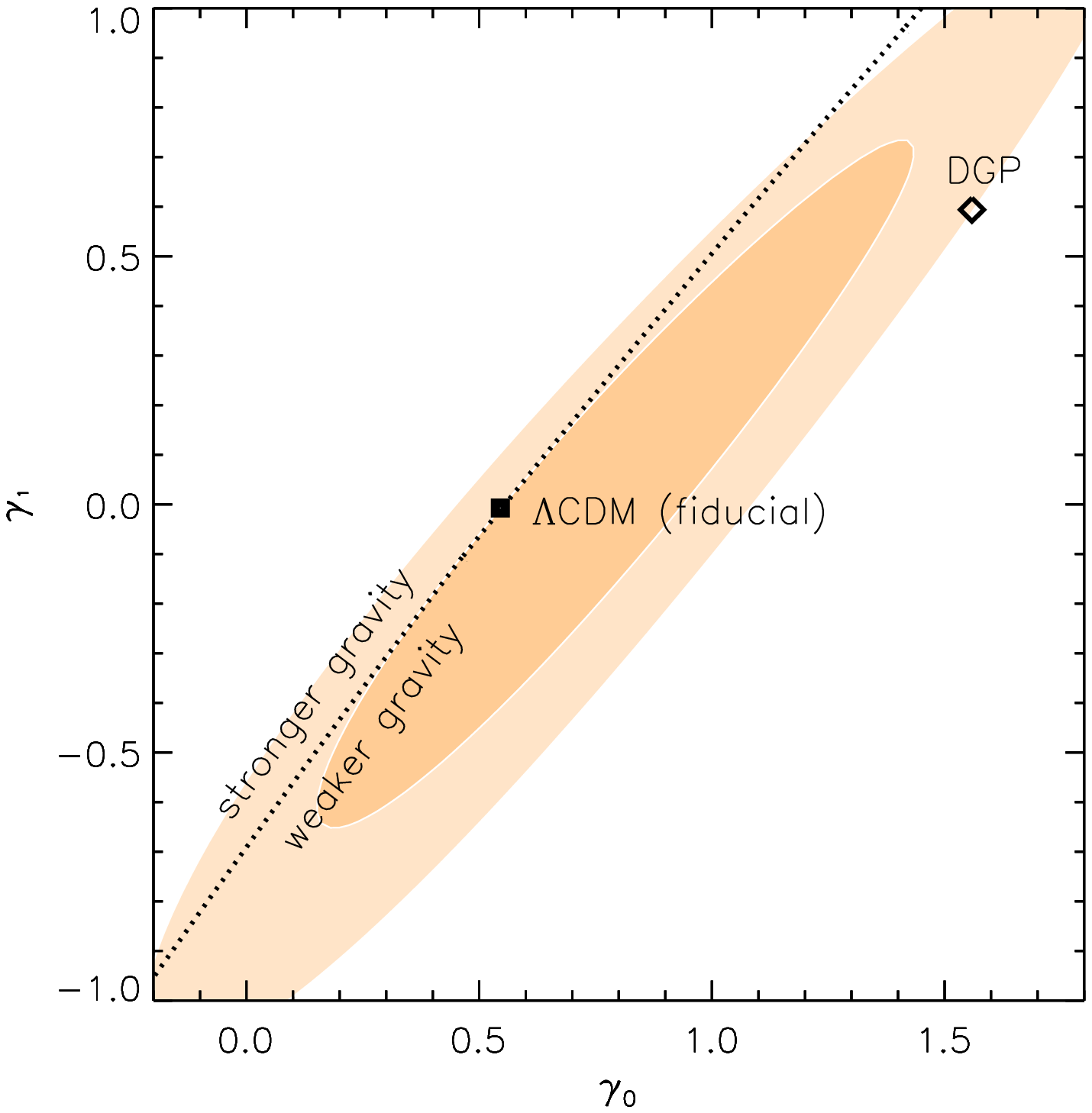}
\includegraphics[scale=0.5]{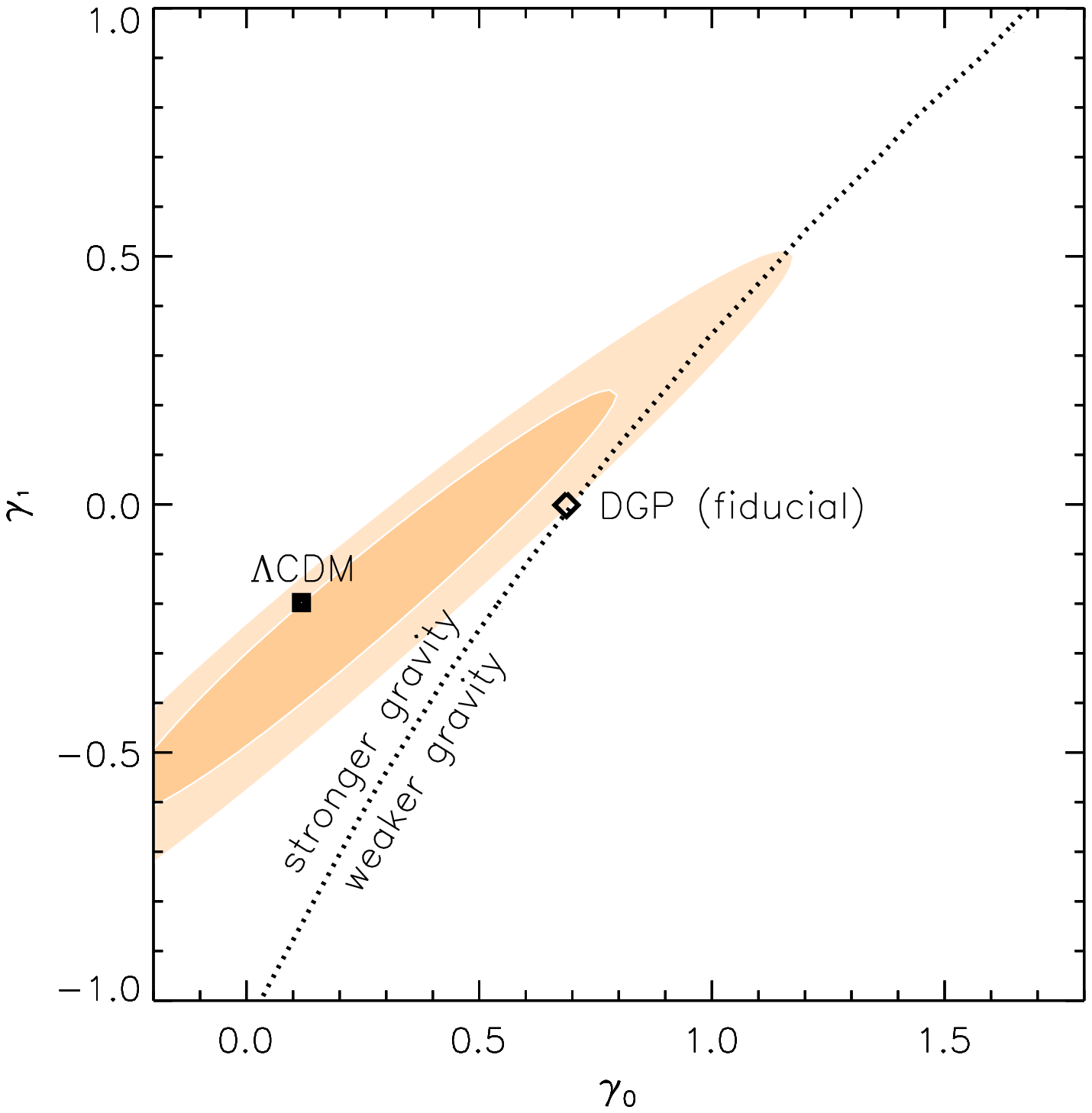}
\caption{{\it Left panel:} $68\%$ and $95\%$ confidence levels in the parameter plane $\gamma_0- \gamma_1$ obtained from  the compilation of data shown in Table \ref{tab:data} and by  using,  as fiducial for the evolution of the background metric,  the expansion rate of the {\it reference} $\Lambda$CDM model ($\Omo=0.315$ and $\sigma_{8,0}=0.835$).  The growth indices theoretically expected in  the fiducial $\Lambda$CDM model is shown by the filled black square.  The growth index expected in the flat DGP model (black diamond) which best fits the expansion rate of the {\it reference} $\Lambda$CDM model  is obtained via \eqref{eq:f_approximation2}, that is after mapping \eqref{gamdgp} in the appropriate background model. {\it Right panel:} The same as the left panel but using as fiducial the expansion rate of the DGP model \cite{DGP} instead.}
\label{fig:lcdm_dgp}
\end{figure}

\begin{figure}[t]
\centering
\includegraphics[scale=0.8]{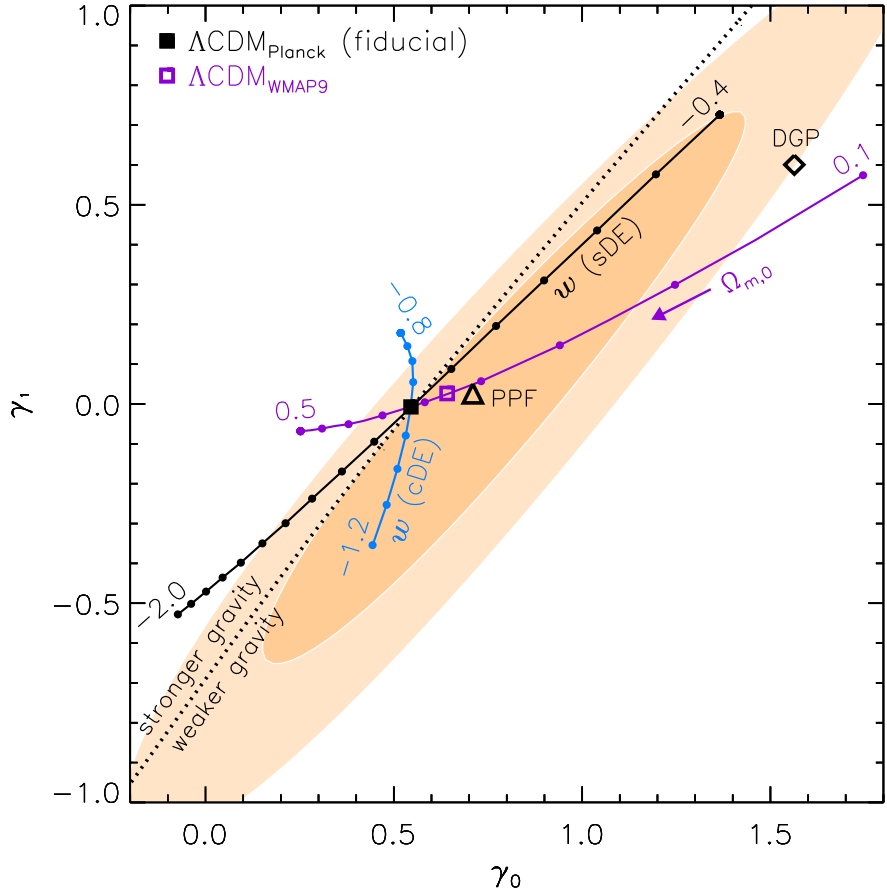}
\caption{Same as in Figure \ref{fig:lcdm_dgp},  but now,  in addition,  we show predictions for different values of $\Omo$ (violet line), the constant $EoS$ for the smooth DE (sDE) model (black line) and the clustering DE (cDE) model (light blue line). The parameter ranges are $0.1 < \Omo <0.5$, $-2.0 < w < -0.4$ (sDE) and $-1.2 < w < -0.8$ (cDE). The spacing is 0.05 for $\Omo$, 0.1 for sDE and 0.05 for cDE. The prediction for the WMAP9 $\Lambda$CDM model are shown by a violet empty square and the black triangle corresponds to a $\zeta$CDM  model with ($w=-1$, $\zeta_1 = 0.6$, $\zeta_2=0$).}
\label{fig:all}
\end{figure}

Up to this point, we have shown  that models with distinct background evolutions and models with distinct growth rate predictions can be analyzed in the same parameter space, thanks to 
the remapping scheme of Eq.~\eqref{eq:f_approximation2}.  
A neat  way to demonstrate the precision and consistency of this strategy is by  showing that the conclusions on the physical viability of a model  are invariant with respect to the choice
of the fiducial in which data are analyzed. 
We will demonstrate this key  feature by exploiting  the DGP\cite{DGP} model which predicts both   different background  and different growth rate evolutions with respect to  the $\Lambda$CDM model.
To this purpose, as in \S \ref{sec:DGP} we consider  the flat DGP model which best fits the expansion rate of the {\it reference} $\Lambda$CDM model. This DGP model has parameters $\mathbf{p}=(0.835,0.213,w_{\rm DGP})$, where the effective DGP $EoS$ is given  in \S \ref{sec:DGP}. 

In Figure \ref{fig:lcdm_dgp} we show  the constraints set on the DGP model by current data (cf.~Table \ref{tab:data}). 
In the left panel, we show the likelihood contours in the parameter plane $\gamma_0 - \gamma_1$ obtained by  using  as fiducial for the evolution of the background metric the expansion rate of the {\it reference} $\Lambda$CDM model. The prediction for the DGP model obtained with the mapping \eqref{eq:f_approximation2} is also shown. 
In the right panel, instead, we show the likelihood contours in the parameter plane $\gamma_0 - \gamma_1$ obtained by  using   the expansion rate of the DGP model as fiducial for the evolution of the background metric. In this case, it is the predictions of the $\Lambda$CDM model that are remapped via Eq.~\eqref{eq:f_approximation2}.  
By confronting these two plots we see that while the location, the amplitude and the orientation of the likelihood surfaces do vary,  
the interpretation of the results is clearly unchanged. Both figures show, consistently, that  measurements of the growth rate in the range $0<z<0.8$ have already a sufficient precision to rule out, at a $95\%$ c.l.,  a most extreme scenario of modified gravity such as the DGP model.  In other terms,  the structure of the likelihood contours depends on the specific fiducial model of the background, that is  
 each fiducial background defines its own growth index parameter space.  But physical interpretation is unchanged if we change from one parameter space to the other using 
 the transformation equation  ~\eqref{eq:f_approximation2}.

The advantage  of the  method is also illustrated  in  Figure \ref{fig:all} using current data. In this picture we simultaneously contrast 
predictions from a large class of  cosmological models against the  likelihood contours derived assuming  the {\it reference} $\Lambda$CDM as fiducial  for data analysis. On top of the DGP model (black diamond), we also plot the effective growth indices  associated to the $\Lambda$CDM model which best fits  WMAP9 data \cite{wmap9} (purple empty square). In this scenario, characterized by the reduced density parameter $\Omo=0.279$, the growth of structures is slightly suppressed with respect to the fiducial, i.e.~the {\it reference} $\Lambda$CDM model of Planck. 
One might also remark the slight (but statistically insignificant) tendency of growth data to favor the WMAP9  substantiation of the $\Lambda$CDM model with respect to the Planck one.
This is a particular  example of a general feature  of $\Lambda$CDM models: the higher the matter  density and the more severe the tension with growth index constraints.
For example $\Lambda$CDM models with $\Omo>0.38$ are excluded at $95\%$ confidence level. 
In Figure  \ref{fig:all} we also show the constraints on DE models with constant  $EoS$  parameter $w$. Smooth DE models with $w<-1.5$  and Clustering DE models with $w>-0.9$ are excluded at $95\%$ confidence level.  Finally, as a curiosity,  we show that a $\zeta$CDM  model (black empty triangle) with the same background as the {\it reference} $\Lambda$CDM model but with a slip parameter $\zeta_1 = 0.6$   best fits  observations, that is it maximizes the likelihood of current data.

\section{Conclusions}
\label{sec:conclusions}
The observational  information about the  growth rate history $f(t)$ of linear cosmic structures can be satisfactorily encoded  into a small set of parameters, 
the growth indices $\gamma_i$, whose amplitude can be analytically predicted by theory. 
Their measurement  allows to explore whether Einstein's field equations encompass gravity also in the infrared, i.e.~on very large  cosmic scales.
In order for this  to be accomplished, {\it a}) an optimal scheme for compressing the growth rate function into the smallest possible set of discrete scalars $\gamma_i$,
without sacrifying  accuracy, and  {\it b})   a prescription for routinely calculating their amplitude in relevant theories of gravity,  in order to  explore the  largest  
region in the space of all possible models, must be devised.

In this paper we have explored a promising  approach towards this goal. We have demonstrated both the precision and the flexibility of a specific parameterization of the growth 
index, that is the logarithmic  expansion \eqref{eq:gamma_Taylor}. 
If the fiducial gravitational  model is not too different from standard GR, i.e.~possible deviations in both the background and perturbed sector 
can be interpreted as first order correction to the Friedmann model, then  the proposed  parameterization scheme 
allows to match numerical results on the redshift dependence of the growth 
index with a relative error which is lower than the nominal precision with which the next generation of redshift surveys are expected to  fix the scaling of this function.
The performances are demonstrated by comparing, for various fiducial gravitational models,  the accuracy of our proposal 
against that of different parameterizations  available in the literature.

Besides accuracy, the formalism features two other critical merits,  one practical and one conceptual.  
First we supply  a simple way for routinely calculating the amplitude of the growth indices in any gravitational model in which the master equation for the 
growth of density perturbations reduces to the form of Eq. \ref{eq:matter_density_fluctuations}. To this purpose it is enough
to specify three characteristic functions of this equation, the expansion rate $H(t)$,  the damping  $\nu(t)$ and the response $\mu(t)$ coefficients to calculate the parameters $\gamma_i$ 
up to any desired order $i$.  Moreover, since the parameterization of the growth rate has not a phenomenological nature, but it is constructed  as a series expansion 
of the exact solutions  of the differential equation which rules the growth of structures (cf. Eq. \ref{eq:f_H_General}), one can easily interpret empirical results about the amplitude 
of the growth indices in terms of fundamental gravitational models. 

Since the growth index is a model dependent quantity, it has been traditionally used only to  reject, statistically, the specific
model adopted to analyze growth data.  We have shown, instead, that the growth index parameter space $\gamma_0-\gamma_1$  provides a diagnostic tool to discriminate a large class of 
models, even those presenting background evolution histories different from the fiducial model adopted in data analysis.  
In other terms,  a detection of a present day growth index amplitude $\neq  0.55$
would not  only indicate a  deviation from $\Lambda$CDM predictions  but could  be used to disentangle  among different alternative explanations of the cosmic acceleration in a straightforward way.
The key to this feature is the mapping of Eq.~\ref{eq:f_approximation2} which allows to  factor out the effect of expansion from the analysis of growth rate histories.   As the standard $\Omo-\Omega_{\Lambda,0}$ plane identifies different expansion histories $H(t)$,  the $\gamma_0-\gamma_1$ plane can thus be used to locate different growth rate histories $f(t)$. 

We have  illustrated  the performance  of the growth index plane in relation to modify gravity model selection/exclusion by 
using current data as well as  forecasts from future experiments. We have shown that the  likelihood contours in the growth index plane $\gamma_0 - \gamma_1$ can be
used to tell apart  a clustering quintessence component \cite{SefVer11}  from a smooth dark energy fluid,  to fix the parameters of  viable Parameterized Post  Friedman gravitational models 
\cite{FerSko10} or  to exclude specific  gravitational models such as, for example,  DGP \cite{DGP}.

The performances  of the analysis tool presented in this paper are expected to be  enhanced, should  the formalism be  coupled to models 
parameterizing  the  large class of possible gravitational  alternatives to standard GR available in the literature. 
In particular various approaches have been proposed to synthetically describe all the possible gravitational laws generated by  adding a single scalar degree of freedom to Einstein's equations 
~\cite{GPV,JLPV, BFS, BFPW}.  Besides   quintessence, scalar-tensor theory
and  $f(R)$ gravity, this formalism allows also to describe covariant Galileons~\cite{NRT}, kinetic gravity braiding \cite{deffa1} and  Horndeski/generalized Galileons theories~\cite{hor,Deffayet:2009wt}. 
Interestingly, the cosmological  perturbation theory  of this general class of models can be 
parameterized so that a direct correspondence between the parameterization  and the underlying space of  theories is maintained.
In a different paper \cite{PSM} we have already explored how the effective field theory  formalism of \cite{GPV} allows to  interpret  the empirical constraints on $\gamma_i$ 
directly in terms of  fundamental gravity theories.

\section*{Acknowledgments}

We acknowledge useful discussions with  L. Guzzo, F. Piazza, T. Sch\"ucker, P. Taxil and J. M. Virey.  CM is grateful for support from specific project funding of the {\it Institut Universitaire
de France} and of the Labex OCEVU.

\section{Appendix A: Amplitude of the growth indices  $\gamma_n$}
\label{sec:Annexe}
We compute the amplitude of the growth indices as follows. First we introduce Eq.~\eqref{eq:gamma_Taylor} into Eq.~\eqref{eq:f_H_x} and obtain
\begin{align}
 \x '(\x ) \Big( \gamma_0 + \sum_{n=1}^{N}(1+n) \gamma_n \frac{\x ^n}{n!}  \Big) + \exp \Big( \sum_{n=0}^{N} \gamma_n \frac{\x ^{n+1}}{n!} \Big) + 1 + \nu(\x ) + \frac{H'}{H}(\x ) \nonumber \\
  - \frac{3}{2} \mu (\x ) \exp \Big(\x  - \sum_{n=0}^{N} \gamma_n \frac{\x ^n+1}{n!} \Big) = \mathcal{O}(\x ^{N+1})
\label{eq:f_H_x_approx}
\end{align}
Neglecting the term on the right hand side, we have transformed the first order non-linear differential equation (\ref{eq:f_H_General}) into an algebraic equation of the form 
\begin{align}
F(\x ) = 0.
\label{eq:F=0}
\end{align}
Now we set $\x =0$ in Eq.~\eqref{eq:f_H_x_approx} and we obtain 
\begin{align}
F(0) = \gamma_0 \:\! \x '(0)  +2 + \nu(0) + \tfrac{H'}{H}(0)  - \tfrac{3}{2} \mu(0) = 0.
\label{eq:F_0}
\end{align}
We suppose that $\x '(0) = 0$, $\tfrac{H'}{H}(0) = -\frac{3}{2}$ and $\mu(0) = \nu(0) =1$ in any viable model.  In order  to compute the coefficients $\gamma_n$ up to an order $N$, we have to expand $F(\x )$ as a Taylor series of $x$ at $\x =0$ up to an order $N+1$:
\begin{align}
F(\x ) = F(0) + \sum_{m=1}^{N+1}\frac{\x ^m}{m!}\Big[\frac{d^{\:\! m}F}{d\x ^m}\Big]_{\x =0} + \mathcal{O}(\x ^{N\:\!\! +\:\!\! 1}).
\label{eq:F_Taylor}
\end{align}
Eq.~\eqref{eq:F=0} is satified if and only if all the coefficients  vanish:
\begin{align}
\Big[\frac{d^{\:\! m}F}{d\x ^{m}}\Big]_{\x =0} = 0 \;, \qquad m=1,2,3,\ldots ,N+1
\label{eq:dF_0}
\end{align}
This leads naturally to the definition of the structural coefficients \eqref{eq:struct}. For $m=1$ in Eq.~\eqref{eq:dF_0} we obtain Eq. \eqref{eq:gamma_0}. Finally, inverting Eq.~\eqref{eq:dF_0} for $m=n+1$ leads to the recursion formula \eqref{eq:gamma_n}.

\end{document}